\documentclass[twocolumn,twocolappendix]{aastex631}

\usepackage{amsthm,amsmath,amssymb}
\usepackage{mathrsfs,mathtools}
\usepackage{graphicx,float}

\shorttitle{Inclination Angles in AGNs}
\shortauthors{Du et al.}

\begin{document}

\title{On the Relation between the Inclination Angle of the Accretion Disk and the Broad-line Region in Active Galactic Nuclei}

\author[0009-0006-6543-6333]{Rong Du}
\affiliation{Department of Astronomy, School of Physics, Peking University, Beijing 100871, China}
\affiliation{Kavli Institute for Astronomy and Astrophysics, Peking University, Beijing 100871, China}

\author[0000-0001-6947-5846]{Luis C. Ho}
\affiliation{Department of Astronomy, School of Physics, Peking University, Beijing 100871, China}
\affiliation{Kavli Institute for Astronomy and Astrophysics, Peking University, Beijing 100871, China}

\author[0000-0002-5770-2666]{Yuanze Ding}
\affiliation{Division of Physics, Mathematics and Astronomy, California Institute of Technology, Pasadena, CA 91125, USA}

\author[0000-0001-8496-4162]{Ruancun Li}
\affiliation{Department of Astronomy, School of Physics, Peking University, Beijing 100871, China}
\affiliation{Kavli Institute for Astronomy and Astrophysics, Peking University, Beijing 100871, China}

\correspondingauthor{Rong Du}
\email{durong@alumni.pku.edu.cn}

\begin{abstract}

Models of active galactic nuclei often invoke a close physical association between the broad-line region and the accretion disk. We evaluate this theoretical expectation by investigating the relationship between the inclination angle of the BLR ($\theta_\mathrm{BLR}$) and the inclination angle of the inner accretion disk ($\theta_\mathrm{disk}$). For a sample of eight active galactic nuclei that have published values of $\theta_\mathrm{BLR}$ estimated from dynamical modeling of the BLR based on velocity-resolved reverberation mapping experiments, we analyze high-quality, joint XMM-Newton and NuSTAR X-ray observations to derive new, robust measurements of $\theta_\mathrm{disk}$ through broadband (0.3--78\,keV) reflection spectroscopy. We find a strong, positive correlation between $\theta_\mathrm{BLR}$ and $\theta_\mathrm{disk}$ (Pearson correlation coefficient 0.856, $p$-value 0.007), although Monte Carlo simulations indicate that the level of significance is only marginal ($<3\,\sigma$). Nevertheless, the nearly linear relation between $\theta_\mathrm{BLR}$ and $\theta_\mathrm{disk}$ suggests of a possible physical alignment between the accretion disk and the BLR.  Future studies with a larger and more homogeneous sample are needed to confirm the correlation and refine our understanding of the structure and dynamics of the central regions of active galaxies.
\end{abstract}


\section{Introduction} \label{sec:intro}

Active galactic nuclei (AGNs) are characterized by the accretion of matter onto their central supermassive black hole (BH) at sub-pc scales, where complex dynamics and multiple physical structures coexist, including the accretion disk and the broad-line region (BLR). The strong broad emission lines arise from BLR gas largely photoionized by intense ultraviolet (UV) to X-ray radiation from the accretion disk \citep[e.g.,][]{Krolik 1998}. Various attempts have been made to model the BLR in the context of accretion disk winds. \cite{Emmering et al. 1992} proposed that clouds in a centrifugally driven magnetohydrodynamic wind extract angular momentum from the accretion disk and are photoionized by the UV radiation from the inner accretion disk. \cite{Murray and Chiang 1995} and \cite{Murray et al. 1995} suggested that the BLR is produced by a smooth, nonspherical disk wind driven by UV emission lines and gas pressure. By contrast, the failed dusty outflow model (e.g., \citealt{Czerny and Hryniewicz 2011, Czerny et al. 2017, Baskin and Laor 2018, Pandey et al. 2023}) claims that the BLR originates from a dust wind in the atmosphere of the disk powered by local radiation pressure; once lifted above the disk, the dust evaporates due to the central emission of the AGN, leading to subsequent fallback of the wind toward the disk. The quasar rain model \citep{Elvis 2000, Elvis 2017}, on the other hand, argues that the BLR arises from a condensed warm disk wind driven by varying radiation pressure, and that, once formed, such clouds rain back down toward the BH and then are destroyed by the warm absorbers (WAs; \citealt{Halpern1984, Reynolds1997}). The disappearance of broad emission lines at very low luminosities \citep{Ho2008} finds a natural explanation in the context of disk wind model for the origin of the BLR \citep[e.g.,][]{Czerny et al. 2004,Elitzur and Ho 2009, Elitzur2014}.

A common feature shared qualitatively among all these models is that BLR has an axisymmetric geometry with an opening angle and an inclination angle, representing the poloidal average of the spatial configuration of the system. The inclination angle of the BLR to the line-of-sight ($\theta_\mathrm{BLR}$) should be roughly similar to the inclination of the disk ($\theta_\mathrm{disk}$). To date, however, this basic prediction has yet to be verified, owing to the observational challenge of constraining the sub-structures of AGNs on such small scales.

Spatially resolving the BLR remains out of reach except for a handful of the brightest AGNs amenable to near-infrared interferometry (e.g., \citealt{GRAVITY Collaboration et al. 2018}). Under most circumstances, the kinematics and structure of the BLR can only be probed indirectly through reverberation mapping \citep[RM;][]{Bahcall et al 1972, Blandford and McKee 1982}, by measuring the time delay between the variations in accretion disk emission and the corresponding changes in the broad emission lines. However, it is nontrivial to interpret the complex information encoded in the RM transfer function \citep{Horne 1994,Skielboe et al. 2015}. Parameterized phenomenological models \citep{Brewer et al. 2011, Pancoast et al. 2011, Pancoast et al. 2012, Pancoast et al. 2014a, Li et al. 2013} have been applied successfully to numerous RM campaigns to extract physical properties of the BLR \citep[e.g.,][]{Bentz et al. 2009, Denney et al. 2009, Denney et al. 2010, Barth et al. 2011a, Barth et al. 2011b, Grier et al. 2013, Du et al. 2016, Du et al. 2018, Pei et al. 2017, De Rosa et al. 2018, Feng et al. 2021, Li et al. 2021}. Such dynamical modeling of velocity-resolved RM data depicts the BLR as an axisymmetric thick geometry with low to moderate $\theta_\mathrm{BLR}$, whose kinematics are largely governed by a combination of Keplerian rotation and some degree of radial motions \citep[e.g.,][]{Pancoast et al. 2014b, Grier et al. 2017, Li et al. 2018, Williams et al. 2018, Williams et al. 2020, Bentz et al. 2021, Villafana et al. 2022}.

Probing structures on the scale of the accretion disk presents even more formidable challenges. Fitting broadened emission lines with a relativistic accretion disk model \citep{Novikov and Thorne 1973, Shakura and Sunyaev 1973}, we can measure $\theta_\mathrm{disk}$ using the double-peaked broad H$\alpha$ emission line \citep[e.g.,][]{Eracleous and Halpern 1994,Storchi-Bergmann et al. 1997,Ho et al. 2000} or the broad Fe~K$\alpha$ 6.4\,keV line when a specific corona configuration is assumed \citep[e.g.,][]{Nandra et al. 1997}. Indirect inferences on $\theta_\mathrm{disk}$ can also be made, for instance by assuming that the disk is normal to the jet and estimating the jet orientation from the core dominance of radio-loud sources \citep[e.g.,][]{Ghisellini et al. 1993,Wills and Brotherton 1995}. In the context of the AGN unified model \citep{Antonucci 1993}, one can attempt a first-order approximation of $\theta_\mathrm{disk}$ by mapping the dynamics of the narrow-line region \citep{Fischer et al. 2013}, or by modeling the infrared spectral energy distribution to constrain the inclination angle of the dusty torus \citep{Zhuang2018}.

\cite{Du et al. 2024} recently tested the  methodology to measure the inclination angle of the inner accretion disk through X-ray reflection spectroscopy. The X-ray emission in AGNs is explained conventionally by hot coronal electrons near the BH inverse-Comptonizing thermal optical or UV photons from the accretion disk into an exponential cutoff power-law continuum \citep[e.g.,][]{Haardt and Maraschi 1991}. Two reflection components often overlie on the power-law continuum: the Fe~K$\alpha$ fluorescence line at $6-7\,\rm keV$ \citep{Fabian et al. 1989} and the Compton scattering hump of the continuum emission peaking at $20-40\,\rm keV$ \citep{Guilbert and Rees 1988}. The corona irradiates the inner disk region, where gravitational redshift, relativistic beaming, and the Doppler effect broaden the Fe~K$\alpha$ line by stretching the low-energy wing and creating a sharper blueshifted peak. The line broadening effects encode information on the velocity of the emitting fluid element as a function of disk inclination and the radius of the emitters, and we can infer $\theta_\mathrm{disk}$ by fitting a broadband spectrum with a numerical reflection model. Advanced models have been developed for calculating a relativistically blurred, broadband reflection continuum, including the \textsc{KY} package \citep{Dovciak et al. 2004}, \textsc{REFLKERR} \citep{Niedzwiecki et al. 2019}, \textsc{RELTRANS} \citep{Ingram et al. 2019}, and \textsc{RELXILL} \citep{Dauser et al. 2014,Garcia et al. 2014}. For our purposes, we adopt the self-consistent, angle-dependent reflection model \textsc{RELXILL}.

Much of the recent interest in BLR modeling is motivated by the desire to measure BH masses in AGNs. So-called single-epoch virial mass estimates of the BH (e.g., \citealt{Vestergaard and Peterson 2006, Ho and Kim 2015}) depend linearly on a scale factor $f_\mathrm{BLR}$, which accounts for the detailed structure, dynamics, and orientation of the BLR. The unknown geometry and inclination angle of the BLR, in particular, lead to larger uncertainties in the $M_\mathrm{BH}$ estimation \citep{Krolik 2001}. While dynamical modeling offers an avenue to estimate $\theta_\mathrm{BLR}$, how can we test the reliability of these results? Verifying that $\theta_\mathrm{BLR} \approx \theta_\mathrm{disk}$ would lend confidence that the BLR inclinations are not too far off the mark. Moreover, a close association between $\theta_\mathrm{BLR}$ and $\theta_\mathrm{disk}$ would also help validate theoretical models that associate the BLR clouds with material in the accretion disk or closely aligned with it. Of course, $\theta_\mathrm{BLR}$ may not align perfectly with $\theta_\mathrm{disk}$. Radiation pressure, magnetic fields, and turbulence can break the symmetry of the system and cause the BLR to be tilted or warped relative to the accretion disk. For instance, the BLR can assume a more spherical distribution in the above-mentioned wind models \citep[e.g.,][]{Emmering et al. 1992,Czerny et al. 2016}, potentially offering insights on AGN feeding and feedback processes. Additional substructures or dynamical interactions in the AGN environment may further complicate any simple expectations of co-planarity between the BLR and the accretion disk.

We aim to study the inclination angles in AGNs, exploring the connection between the accretion disk and other facets of the AGN and its host galaxy. We investigate the possible correlation between the inclination angle of the BLR and the inclination angle of the inner accretion disk in a small sample of nearby AGNs that have both $\theta_\mathrm{BLR}$ measurements secured from published RM dynamical modeling and $\theta_\mathrm{disk}$ derived in this study through X-ray reflection spectroscopy. We adopt a cosmology with $H_0 = 70\, \rm km\,s^{-1}\,Mpc^{-1}$, $\Omega_\Lambda = 0.73$, and $\Omega_m = 0.27$.

\section{Experimental Design and Methodology}
\subsection{Sample Definition} \label{sec:sample}
Our objective is to derive $\theta_\mathrm{disk}$ measurements for AGNs that already have published values of $\theta_\mathrm{BLR}$. To this end, we start by assembling all type~1 (broad-line) AGNs with high-quality RM observations for which phenomenogical dynamical models of the BLR have been constructed using the Code for AGN Reverberation and Modeling of Emission Lines (\textsc{CARAMEL}) developed by \cite{Pancoast et al. 2011, Pancoast et al. 2014a}. The model has been applied systematically to 28 AGNs with velocity-resolved RM data of the H$\beta$ emission line \citep{Pancoast et al. 2014b, Grier et al. 2017, Williams et al. 2018,Williams et al. 2020, Bentz et al. 2021,Bentz et al. 2022, Villafana et al. 2022,Villafana et al. 2023}. These form our primary sample.

Additionally, we acknowledge the Bayesian estimator developed by \cite{Li et al. 2013, Li et al. 2018}, which constrains the size and structure of the BLR through dynamical modeling based on \cite{Pancoast et al. 2011}. This method incorporates the nonlinear emission response of the BLR to continuum variations and has been successfully applied to several AGNs \citep[e.g.,][]{Du et al. 2014, Du et al. 2015, Du et al. 2016, Du et al. 2018, Wang et al. 2014, Li et al. 2018, Xiao et al. 2018, Lu et al. 2019}. However, we choose not to include results from this approach in the present study to avoid potential inconsistencies arising from different BLR modeling techniques. Moreover, most AGNs in their sample are super-Eddington sources, which do not overlap with our primary sample. These sources may also introduce conflicts with our X-ray reflection modeling, which is based on the assumption of a standard accretion disk.

For our purpose, the X-ray observations must satisfy a number of requirements in order to obtain reliable measurements of $\theta_\mathrm{disk}$. Guided by the tests presented in \cite{Du et al. 2024}, we search for multi-epoch, high-quality X-ray data from the public data archives of the X-ray Multi-Mirror Mission \citep[XMM-Newton;][0.3--10\,keV]{Jansen et al. 2001} and the Nuclear Spectroscopy Telescope Array \citep[NuSTAR;][3--78\,keV]{Harrison et al. 2013}. We favor sources with simple absorbing features to reduce the complexity of the spectral fitting and to minimize the uncertainties in $\theta_\mathrm{disk}$ caused by this contaminant. From the original sample of 28 AGNs with RM-based measurements of $\theta_\mathrm{BLR}$, only eight sources remain that have clearly measured reflection components yet with simply modeled WAs. Table~\ref{tab:observations} summarizes the X-ray observations of the final sample, among which \cite{Williams et al. 2018} analyzed Mrk\,279, Mrk\,50, PG\,1310$-$108, and Zw\,229$-$015 as part of the Lick AGN Monitoring Project (LAMP) 2011 \citep{Barth et al. 2015} and \cite{Villafana et al. 2022} analyzed MCG\,+04$-$22$-$042, Mrk\,1392, PG\,2209+184, and RBS\,1917 in connection with the LAMP 2016 program \citep{U et al. 2022}. 

\subsection{Derivation of Accretion Disk Inclination} \label{sec:fit_method}
The inner accretion disk inclination, formally defined as the viewing angle with respect to its norm, is measured by jointly fitting the broadband X-ray spectra of AGNs using the self-consistent, angle-dependent reflection model \textsc{RELXILL} \citep{Dauser et al. 2014,Garcia et al. 2014}. This model solves for the inclination $\theta_\mathrm{disk}$, dimensionless spin parameter $a_\ast$, iron abundance $A_\mathrm{Fe}$, power-law photon index $\Gamma$, ionization state $\xi$, electron temperature $kT_e$, reflection fraction $R_f$, and coronal emissivity parameters Index$_\mathrm{1}$ and Index$_\mathrm{2}$, and the broken radius $R_\mathrm{br}$. Specifically, we use \textsc{RELXILLCP} in the \textsc{RELXILL} family to account for the soft excess, broad Fe~line, and Compton hump \citep[see Section~5.6 of][]{Du et al. 2024}.

We use spectra retrieved from the PN charge-coupled device and Metal Oxide Semiconductor (MOS) of the European Photon Imaging Camera (EPIC) onboard XMM-Newton, and from the two focal plane modules (FPMA/B) of NuSTAR. We select the 0.3--10\,keV band for XMM-Newton and the 3--78\,keV band for NuSTAR, ignoring all bad spectral bins. We divide observations chronologically into different epochs for each source and fit the epochs jointly. Spectral fittings are conducted using \textsc{Xspec} \citep[12.12.1;][]{Arnaud 1996} with the modified \cite{Levenberg 1944}--\cite{Marquardt 1963} algorithm to minimize the $\chi^2$ statistic \citep{Bevington 1969}, and posterior parameter distributions are generated to determine confidence intervals.

\begin{deluxetable*}{cccccc}
    \label{tab:observations}
    \tablenum{1}
    \caption{Summary of X-ray Observations}
    \tablehead{
        \colhead{Name} &
        \colhead{Redshift\tablenotemark{a}} &
        \colhead{Epoch} &
        \colhead{Observatory} &
        \colhead{Observation ID} &
        \colhead{Effective Exposure (ks)}\\
        \colhead{(1)} & \colhead{(2)} & \colhead{(3)} & \colhead{(4)} & \colhead{(5)} & \colhead{(6)}
    }
    \startdata
MCG\,+04$-$22$-$042    &0.03311 & $a$ & XMM-Newton & 0312191401 & 1.4 (PN) / 9.2 (MOS) \\ 
                       &        & $b$ & NuSTAR & 60061092002 & 18.8 \\ 
                       &        & $c$ & NuSTAR & 60602018002 & 43.0 \\ 
                       &        & $d$ & NuSTAR & 60602018004 & 42.6 \\ 
                       &        & $e$ & NuSTAR & 60602018006 & 40.7 \\ \hline
Mrk\,50                &0.02386 & $a$ & XMM-Newton & 0650590401 & 11.7 \\ 
                       &        & $b$ & NuSTAR & 60061227002 & 17.4 \\ \hline
Mrk\,279               &0.03045 & $a$ & XMM-Newton & 0872391301 & 20.0 \\ 
                       &        & $b$ & NuSTAR & 60601011004 & 200.6 \\ \hline
Mrk\,1392              &0.03588 & $a$ & XMM-Newton & 0795670101 & 25.6 \\ 
                       &        & $b$ & NuSTAR & 60160605002 &  21.1 \\ \hline
PG\,1310$-$108         &0.03427 & $a$ & XMM-Newton & 0801891601 & 19.4 \\ \hline
PG\,2209$+$184         &0.06990 & $a$ & \begin{tabular}[c]{@{}c@{}}XMM-Newton \cr NuSTAR\end{tabular} & \begin{tabular}[c]{@{}c@{}}0795620201 \cr 60301015002 \\ \end{tabular} & \begin{tabular}[c]{@{}c@{}}36.2 \cr 101.9 \\ \end{tabular} \\ \hline
RBS\,1917              &0.06600 & $a$ & XMM-Newton & 0762871101 & 33.6 \\ \hline
Zw\,229$-$015          &0.02788 & $a$ & XMM-Newton & 0672530301 & 27.5 \\ 
                       &        & $b$ & NuSTAR & 60160705002 & 22.0
    \enddata
\tablenotetext{a}{Redshift obtained from the NASA/IPAC Extragalactic Database (NED).}
\end{deluxetable*}

\subsection{Observations and Data Reduction} \label{sec:data_reduction}
Data from both XMM-Newton and NuSTAR for the eight sources in our sample are gathered and processed to retrieve the spectra. We divide the spectra of each source into different epochs according to the observed date and respective instrument (Table~\ref{tab:observations}). For XMM-Newton data, light curves and spectra are extracted from EPIC detectors using System Analysis Software (version 20.0.0; calibration files version 31 of March 2022). Event lists are retrieved using the tasks \texttt{epproc} and \texttt{emproc}, source and background spectra are extracted with \texttt{evselect}, and ancillary files and redistribution matrices are generated with \texttt{arfgen} and \texttt{rmfgen}. In the absence of pile-up, source spectra are extracted from a 40\arcsec-diameter circular region. For Mrk\,50, corrections for pile-up are made using an annulus region between 10\arcsec\ and 30\arcsec\ in diameter. Background spectra are extracted from source-free polygon regions for PN observations and a 300\arcsec-diameter circular region for MOS observations. Spectra are rebinned with \texttt{specgroup} to have at least 25 counts per bin and not oversample the full width at half-maximum resolution by more than a factor of 3. For NuSTAR data, event lists from both FPM detectors are retrieved with \texttt{NUPIPELINE} in \textsc{NuSTARDAS} (2.1.2; calibration files from \textsc{CALDB} version 20220802) within the \textsc{HEASoft} bundle (6.30.1). With \texttt{NUPRODUCTS}, spectra are extracted from a 40\arcsec-diameter circular region centered on the source, with background spectra from a source-free 300\arcsec- diameter circle near the source. Spectra are rebinned with \texttt{ftgrouppha} to have at least 25 counts per bin.

\subsection{Spectral Fitting}
\label{sec:fitting}
We analyze the sources with a baseline model that takes into consideration Galactic absorption, blackbody emission, the reflection continuum, and WAs\footnote{In \textsc{Xspec} terminology: \\ \textsc{CONSTANT*TBABS*XSTAR*(ZBBODY+RELXILLCP)}.}. For each source, we initially fit a phenomenological redshifted power-law model (\textsc{ZPOWERLW}) modified by Galactic absorption \citep[\textsc{TBABS}; see][]{Wilms et al. 2000}. Broadband fitting with systematic residuals at soft and hard energies motivate using \textsc{RELXILLCP} to model the soft excess, the broad Fe~K$\alpha$ line, and the high-energy Compton hump. As in \citealp{Du et al. 2024} (see their Section~5.7), we also model the mild soft excess with an additional epoch-independent redshifted blackbody (\textsc{ZBBODY}) component. WAs in the soft band are fit with multiplicative \textsc{XSTAR} tables, with three free parameters per XMM-Newton epoch: hydrogen column density $N_\mathrm{H}$, ionization parameter $\xi_\mathrm{WA}$, and outflow velocity $v$. The \textsc{XSTAR} tables are generated with an input power-law continuum with photon index $\Gamma = \{1.0, 3.0, 5.0, ..., 15.0\} \times 10^{-1}$, turbulent velocity $500\, \rm km~s^{-1}$, $\log (N_\mathrm{H}/\rm cm^{-2}) = \{$19.0, 19.5, 20.0, 20.5, ..., 23.0$\}$, and $\log (\xi_\mathrm{WA}/\mathrm{erg\,cm\,s^{-1}}) = \{-2.0, -2.0 + 5/9, -2.0 + 2 \times 5/9, ... , 3.0\}$. Although multi-phase WAs or ultra-fast outflows are likely to exist in some sources, it is beyond the scope of our analysis to deal with the complicated soft-band emission, absorption, and reprocession features, which, in any case, would require higher resolution soft X-ray spectra, such as those from the Reflection Grating Spectrometer \citep[RGS;][]{den Herder et al. 2001}. One \textsc{XSTAR} table is usually sufficient in this work.

During the fit, the Galactic absorption column density is fixed to values from the Leiden/Argentine/Bonn Galactic H\,{\small I} Survey \citep{Kalberla et al. 2005}. The blackbody temperature $kT$ is a free parameter but assumed constant across spectral epochs. For \textsc{RELXILLCP}, disk-related parameters ($\theta_\mathrm{disk}$, $a_\ast$, and $A_\mathrm{Fe}$) are fixed across all epochs, while corona-related parameters ($\Gamma$, $kT_e$, $R_f$, Index$_\mathrm{1}$, Index$_\mathrm{2}$, $R_\mathrm{br}$, and $\xi$) are free between different epochs. Inner radius ($R_\mathrm{in}=$ radius of the innermost stable circular orbit), outer radius ($R_\mathrm{out} = 400r_{g}$, where $r_{g} \coloneqq GM/c^2$), and density ($n = 10^{15}\,\mathrm{cm^{-3}}$) are fixed. Cross-calibration constants account for instrument differences, and normalization parameters compensate for variations in integrated flux and exposure time.

Upon achieving the best fit, we use a Markov chain Monte Carlo \citep[MCMC;][]{Metropolis et al. 1953} algorithm to generate the posterior probability distributions of the baseline parameters. We use the \cite{Goodman and Weare 2010} algorithm with 200 walkers, generating proposals from a Gaussian distribution around the best-fit values. Chain convergence is tested with integrated autocorrelation time $\tau_{f}$ using \textsc{EMCEE} \citep{Foreman-Mackey et al. 2013}, with reference to the evolution history of the model parameters and fit statistics. Chains run longer than $\sim 1000 \, \tau_{f}$ \citep{Sokal 1996}, with a 10\% burn-in rejection of initial chain elements. We calculate the 90\% confidence intervals of all parameters from the MCMC chains. 

Table~\ref{tab:result_refl} gives an overview of the best-fit values of the continuum. We summarize the best-fit values of the accretion disk inclination, the BLR inclination, together with the galactic disk inclination (see Section~\ref{sec:galactic}) in Table~\ref{tab:angles}. In Appendix~\ref{app:details}, we provide details of the spectral fitting for each source, and the results for the WAs are concluded in Table~\ref{tab:absorbers}.

\movetabledown=20mm
\begin{rotatetable*}
    \centerwidetable
    \begin{deluxetable*}{ccccccccccccc}
        \tablenum{2}
        \tablecaption{Best-fit Parameters for the X-ray Continuum\label{tab:result_refl}}
        \tablehead{
        \colhead{Name} &
        \colhead{Epoch} &
        \colhead{$\theta_\mathrm{disk}$} &
        \colhead{$a_\ast$} &
        \colhead{$A_\mathrm{Fe}$} &
        \colhead{$R_f$} &
        \colhead{$kT_e$} &
        \colhead{$\Gamma$} &
        \colhead{Index$_\mathrm{1}$} &
        \colhead{Index$_\mathrm{2}$} &
        \colhead{$R_\mathrm{br}$} &
        \colhead{$\log \xi$} &
        \colhead{$L_\mathrm{0.1-200\,keV}$}\\
        &&\colhead{($^\circ$)} && \colhead{($A_\odot$)}&& \colhead{(keV)} &&&&\colhead{($r_{g}$)}& \colhead{($\mathrm{erg~cm~s^{-1}}$)} &
        \colhead{($10^{44}\, \mathrm{erg~s^{-1}}$)}\\
        \colhead{(1)} & \colhead{(2)} & \colhead{(3)} & \colhead{(4)} & \colhead{(5)} & \colhead{(6)} & \colhead{(7)} & \colhead{(8)} & \colhead{(9)} & \colhead{(10)} & \colhead{(11)} & \colhead{(12)} &  \colhead{(13)}
        }
        \startdata
        MCG\,+04$-$22$-$042 & $a$ & & & & $2.82_{-0.32}^{+0.50}$ & $91_{-23}^{+15}$ & $1.91 \pm 0.04$ & $8.2_{-0.3}^{+0.4}$ & $1.5_{-0.2}^{+0.3}$ & $7.7_{-1.0}^{+1.1}$ & $2.87_{-0.19}^{+0.13}$ & $0.90 \pm 0.02$ \\
        & $b$ & & & & $0.32_{-0.12}^{+0.11}$ & $72_{-11}^{+10}$ & $1.84_{-0.04}^{+0.05}$ & $9.0_{-1.1}^{+0.8}$ & $2.2 \pm 0.4$ & $3.0_{-0.8}^{+0.9}$ & $2.96_{-0.18}^{+0.08}$ & $2.36 \pm 0.11$ \\
        & $c$ & $15.2_{-1.5}^{+2.2}$ & $0.996_{-0.013}^{+0.002}$ & $ >8.0$ & $0.97_{-0.15}^{+0.18}$ & $20 \pm 4$ & $1.82 \pm 0.04$ & $8.8_{-0.5}^{+0.5}$ & $2.5_{-0.2}^{+0.3}$ & $3.1_{-0.4}^{+0.4}$ & $2.45_{-0.59}^{+0.36}$ & $1.82 \pm 0.06$ \\
        & $d$ & & & & $0.73_{-0.28}^{+0.14}$ & $74_{-18}^{+11}$ & $1.87 \pm 0.04$ & $8.6 \pm 1.1$ & $1.6_{-0.4}^{+0.2}$ & $4.9_{-0.6}^{+1.1}$ & $2.76_{-0.29}^{+0.19}$ & $1.93 \pm 0.06$ \\
        & $e$ & & & & $1.15_{-0.31}^{+0.29}$ & $> 400$ & $1.83_{-0.10}^{+0.07}$ & $8.9_{-1.5}^{+0.9}$ & $2.3 \pm 0.2$ & $3.6_{-0.4}^{+0.6}$ & $3.01_{-0.56}^{+0.31}$ & $1.12 \pm 0.05$ \\ \hline
        Mrk\,50 & $a$ & $19.1_{-3.3}^{+2.0}$ & $0.992_{-0.030}^{+0.005}$ & $3.5_{-0.2}^{+0.4}$ & $2.84_{-0.42}^{+0.37}$ & $>400$ & $1.75_{-0.06}^{+0.05}$ & $>9.5$ & $4.3_{-0.5}^{+0.4}$ & $2.4_{-0.2}^{+0.3}$ & $3.24_{-0.08}^{+0.06}$ & $0.22 \pm 0.01$ \\
        & $b$ & & & & $0.59_{-0.12}^{+0.09}$ & $24_{-3}^{+4}$ & $1.76_{-0.05}^{+0.07}$ & $2.5_{-0.6}^{+0.4}$ & $7.8_{-2.1}^{+1.0}$ & $20.1_{-2.9}^{+2.5}$ & $3.13_{-0.27}^{+0.24}$ & $0.45 \pm 0.04$ \\ \hline
        Mrk\,279 & $a$ & $26.9_{-3.7}^{+2.5}$ & $0.997_{-0.006}^{+0.001}$ & $6.3_{-1.5}^{+2.4}$ & $0.57_{-0.24}^{+0.66}$ & $20_{-9}^{+10}$ & $1.86 \pm 0.01$ & $>7.3$ & $2.3_{-0.2}^{+0.1}$ & $3.1_{-0.2}^{+0.4}$ & $3.14_{-0.10}^{+0.08}$ & $1.13 \pm 0.01$ \\
        & $b$ & & & & $1.90_{-0.73}^{+1.00}$ & $42_{-14}^{+18}$ & $1.81_{-0.04}^{+0.07}$ & $>8.9$ & $1.0_{-0.6}^{+0.5}$ & $5.1_{-0.8}^{+1.3}$ & $1.48_{-0.57}^{+0.39}$ & $0.68 \pm 0.01$ \\ \hline
        Mrk\,1392 & $a$ & $25.5_{-9.9}^{+6.7}$ & $0.942_{-0.093}^{+0.038}$ & $0.51_{-0.01}^{+0.07}$ & $>7.36$ & $257_{-110}^{+124}$ & $2.05 \pm 0.07$ & $6.9_{-1.1}^{+1.7}$ & $3.2 \pm 0.7$ & $4.7_{-1.1}^{+2.3}$ & $2.46_{-0.22}^{+0.20}$ & $0.53 \pm 0.01$ \\
        & $b$ & & & & $4.78_{-1.63}^{+2.11}$ & $170_{-98}^{+100}$ &  $1.82_{-0.09}^{+0.10}$ & $6.7_{-3.0}^{+2.5}$ & $1.4_{-1.2}^{+0.8}$ & $4.8_{-1.7}^{+2.2}$ & $3.26_{-0.18}^{+0.33}$ & $0.72 \pm 0.06$ \\ \hline
        PG\,1310$-$108 & $a$ & $33.5_{-22.4}^{+20.8}$ & $0.903_{-0.403}^{+0.091}$ & $4.7_{-1.5}^{+1.6}$ & $< 2.02$ & $23_{-12}^{+35}$ & $1.90_{-0.09}^{+0.03}$ & $4.4_{-3.4}^{+2.7}$ & $<4.4$ & $39_{-31}^{+75}$ & $3.01_{-0.13}^{+0.58}$ & $0.47 \pm 0.01$ \\ \hline
        PG\,2209+184 & $a$ & $35.6_{-25.9}^{+31.7}$ & $0.989_{-0.079}^{+0.009}$ & $4.7_{-1.4}^{+1.6}$ & $0.42_{-0.21}^{+0.29}$ & $89_{-40}^{+55}$ & $1.90 \pm 0.02$ & $6.6_{-3.7}^{+2.6}$ & $2.1_{-1.6}^{+0.8}$ & $5.1_{-2.3}^{+2.6}$ & $2.92_{-0.15}^{+0.11}$ & $1.78 \pm 0.01$ \\ \hline
        RBS\,1917 & $a$ & $17.2_{-1.1}^{+4.5}$ & $0.977_{-0.003}^{+0.012}$ & $4.1_{-0.1}^{+0.1}$ & $> 9.87$ & $> 380$ & $2.11_{-0.02}^{+0.07}$ & $>8.6$ & $3.1 \pm 0.1$ & $4.5_{-0.7}^{+0.2}$ & $2.68_{-0.18}^{+0.04}$ & $1.63 \pm 0.02$ \\ \hline
        Zw\,229$-$015 & $a$ & $26.2_{-0.5}^{+2.5}$ & $0.993_{-0.059}^{+0.005}$ & $6.3_{-0.1}^{+0.4}$ & $0.50_{-0.03}^{+0.10}$ & $52_{-7}^{+32}$ & $1.78_{-0.02}^{+0.01}$ & $6.5_{-0.1}^{+0.5}$ & $7.8_{-0.3}^{+0.1}$ & $<1.1$ & $2.87 \pm 0.01$ & $0.18 \pm 0.01$ \\
        & $b$ & & & & $1.38_{-0.06}^{+0.27}$ & $14_{-1}^{+2}$ & $1.80_{-0.01}^{+0.10}$ & $3.6_{-0.1}^{+0.1}$ & $8.9_{-0.5}^{+0.1}$ & $2.1_{-0.1}^{+0.4}$ & $1.55_{-0.12}^{+0.02}$ & $0.40 \pm 0.03$
        \enddata
        \tablecomments{
        Col. (1): Name of the source.
        Col. (2): Epoch.
        Col. (3): Inclination angle of the accretion disk.
        Col. (4): Dimensionless BH spin.
        Col. (5): Iron abundance in units of the solar abundance.
        Col. (6): Reflection fraction.
        Col. (7): Temperature of the corona.
        Col. (8): Photon index.
        Col. (9): Power-law index 1 for the broken power-law disk emissivity.
        Col. (10): Power-law index 2 for the broken power-law disk emissivity.
        Col. (11): Break radius for the broken power-law disk emissivity, in units of the gravitational radius $r_{g}=GM/c^{2}$.
        Col. (12): Ionization parameter of the accretion disk.
        Col. (13): Luminosity in the 0.1--200\,keV band of the power-law continuum.
        For sources observed in more than one epoch, the spin, inclination, and iron abundance are fit simultaneously for all epochs. The best-fit values and 90\% confidence intervals are presented.}
    \end{deluxetable*}
\end{rotatetable*}

\begin{deluxetable}{clllcc}
    \tablenum{3}
    \tablecaption{Key Angles of the Eight Sources\label{tab:angles}}
    \tablehead{
        \colhead{Name} &
        \colhead{$\theta_\mathrm{disk}$} &
        \colhead{$\theta_\mathrm{BLR}$} &
        \colhead{$\theta_\mathrm{o}$} &
        \colhead{$q$} &
        \colhead{$\theta_\mathrm{gal}$}
        \\
        &
        \colhead{($^\circ$)} &
        \colhead{($^\circ$)} &
        \colhead{($^\circ$)} &
        &
        \colhead{($^\circ$)}
    }
    \decimalcolnumbers
    \startdata
    MCG\,+04$-$22$-$042 & $15.2_{-1.5}^{+2.2}$ & $11.3_{-5.0}^{+5.8}$ & $13.6_{-4.9}^{+6.9}$ & 0.56 & 57.7 \\
    Mrk\,50     & $19.1_{-3.3}^{+2.0}$ & $19.8_{-5.4}^{+6.0}$ & $14.1_{-3.7}^{+4.8}$ & 0.72 & 45.1 \\
    Mrk\,279 & $26.9_{-3.7}^{+2.5}$ & $29.1_{-3.4}^{+3.4}$ & $41.0_{-4.1}^{+4.3}$ & 0.70 & 46.8 \\
    Mrk\,1392 & $25.5_{-9.9}^{+6.7}$ & $25.5_{-2.8}^{+3.4}$ & $41.2_{-4.8}^{+5.3}$ & 0.50 & 62.1 \\
    PG\,1310$-$108 & $33.5_{-22.4}^{+20.8}$ & $44.0_{-13.0}^{+35.0}$ & $58.0_{-16.0}^{+25.0}$ & 0.86 & 31.4 \\
    PG\,2209+184 & $35.6_{-25.9}^{+31.7}$ & $30.2_{-6.9}^{+8.7}$ & $29.1_{-8.4}^{+11.0}$ & 0.84 & 33.6 \\
    RBS\,1917 & $17.2_{-1.1}^{+4.5}$ & $20.2_{-3.9}^{+9.9}$ & $25.1_{-7.5}^{+9.2}$ & 0.70 & 46.8 \\
    Zw\,229$-$015 & $26.2_{-0.5}^{+2.5}$ & $32.9_{-5.2}^{+6.1}$ & $33.5_{-6.2}^{+6.4}$ & 0.56 & 57.7
    \enddata
    \tablecomments{
        Col. (1): Name of the source.
        Col. (2): Inclination angle of the inner accretion disk with 90\% confidence intervals from our X-ray reflection spectroscopy as detailed in Appendix~\ref{app:details}.
        Col. (3): Inclination angle of the BLR with 68\% confidence intervals collected from \cite{Williams et al. 2018} and \cite{Villafana et al. 2022}.
        Col. (4): Opening angle of the BLR with 68\% confidence intervals also collected from both works.
        Col. (5): Axial ratio $q = b/a$ in the 2MASS $K_s$ band from \cite{Jarrett et al. 2000}.
        Col. (6): Inclination angle of the galactic disk.
    }
\end{deluxetable}

\begin{figure}
    \centering
    \includegraphics[width=1.00\linewidth]{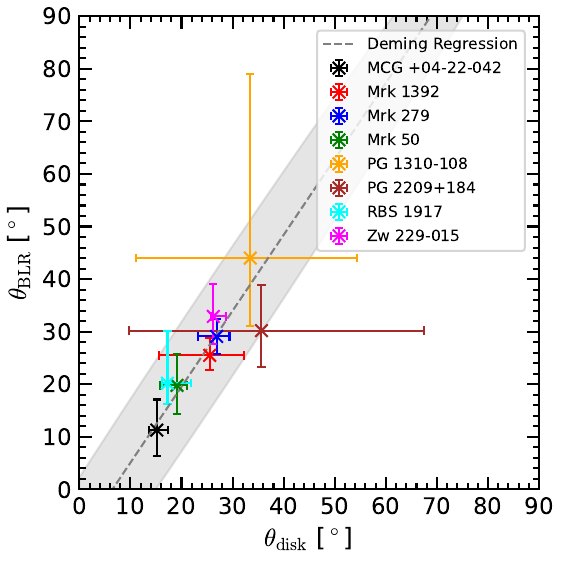}
    \caption{The relation between the BLR inclination versus the inner accretion disk inclination. The gray dashed line depicts the 
    relation obtained from the Deming regression. The gray shaded region represents the 90\% confidence band of the regression.}
    \label{fig:incl_corr}
\end{figure}

\section{Results and Discussion} \label{sec:discussion}

\subsection{Correlation Analysis}

We plot the inclinations of the BLR against those of the inner accretion disk in Figure~\ref{fig:incl_corr}. The standard Pearson correlation coefficient between the two inclinations is 0.856, and a $T$-test with null hypothesis of no correlation returns a $p$-value of 0.007. Therefore, we have $\sim 99\%$ confidence ($\sim 2.72\,\sigma$) that a strong positive relationship exists. We run Monte Carlo simulations to determine the probability that such a one-to-one correlation might arise from an actual underlying relation given the errors of the inclination measurements. We generate random data points for both inclinations with Gaussian distributions of the measured values and their errors, and calculate the correlation coefficient and $p$-value for each simulation. From $10^3$ to $10^6$ realizations, the probability of obtaining a $p$-value less than 0.05 is 39\%, and the probability of obtaining a $p$-value less than 0.01 converges to $\sim 15\%$, with a mean correlation coefficient of 0.56. The significance levels for both the existence and absence of a genuine correlation are at a subtle level of $<3\,\sigma$. Considering the empirically derived errors of both variables, a \cite{Deming 1943} regression gives

\begin{equation}
\theta_\mathrm{BLR} = (1.444 \pm 0.648) \times \theta_\mathrm{disk} + (-9.335 \pm 13.569),
\end{equation}

\noindent
where the 90\% confidence intervals of the slope and intercept are computed with the Jackknife resampling \citep{Quenouille 1949,Quenouille 1956,Tukey 1958}. Notably, the slope of the relation is $\sim 1$, which suggests that the value of $\theta_\mathrm{BLR}$ of a source is expected statistically to be close to its $\theta_\mathrm{disk}$. Nevertheless, a null hypothesis test of the slope being 1 returns a $p$-value of 0.232 (i.e., 77\% confidence), and we stress that, according to our Monte Carlo simulations that take the errors into consideration, the correlation, and hence the significance of the regression, is marginal.

Only eight sources are included in this work. The current analysis clearly suffers from insufficient data and the inhomogeneity of the type~1 AGN sample, which clusters toward inclinations below $50^\circ$. Indeed, we are fortunate that most sources have $\theta_\mathrm{disk}$ around $30^\circ \pm 10^\circ$, which is close to the average value of $\theta_\mathrm{BLR}$, in order to return a somewhat strong positive correlation coefficient close to 1. To illustrate how tenuous the current results are, note that, for instance, if a point with $\theta_\mathrm{BLR} \approx \theta_\mathrm{disk} = 10^\circ$ were added to the current sample, the correlation coefficient would rise to 0.897 and the significance level to $3.28\,\sigma$. Instead, if a point with $\theta_\mathrm{BLR} \approx \theta_\mathrm{disk} = 75^\circ$ point were to be introduced, the correlation coefficient and significance level would be boosted to 0.964 and $4.20\,\sigma$, respectively. A major challenge of this work stems from the fact that we have restricted our analysis to sources that do not experience strong X-ray obscuration, to ensure that the spectral fits can successfully derive the inclinations; otherwise, seldom can we manage to interpret the spectra with a correct model for any measurement. Firmer conclusions on the relationship between BLR and accretion disk inclination must await future analysis with a significantly larger number of sources.

\subsection{Caveats for the Accretion Disk Inclination}

The uncertainties on $\theta_\mathrm{disk}$ span a wide range related to their statistical and systematic aspects.

\subsubsection{Statistical Uncertainties}

In practice, the reliability of $\theta_\mathrm{disk}$ measurements is heavily influenced by factors including the number of spectral epochs available for analysis and the simultaneity of observations from different instruments; the reflection fraction of the source also matters, as do the iron abundance and the coronal temperature \citep{Du et al. 2024}. However, $R_f$ usually stands out as the major factor \citep{Du et al. 2024}. As a result of these complications, directly comparing the error bars on $\theta_\mathrm{disk}$ among the eight sources is hardly instructive. Limiting to the statistical behaviors, we highlight this difficulty with two concrete examples.

With five epochs of nonsimultaneous observations (one from XMM-Newton, four from NuSTAR), MCG\,+04$-$22$-$042 has one of the smallest errors in our sample. In comparison, PG\,1310$-$108, with only one epoch of XMM-Newton observation, has nearly the largest error. Without NuSTAR providing information about the higher energy band, it is hard to model the entire reflection continuum correctly in the absence of the Compton hump \citep[e.g.,][]{Fabian et al. 2015, Du et al. 2024}. Moreover, as per \cite{Du et al. 2024}, $R_f$ can be treated as a qualitative indicator for underlying uncertainties of the spectral fittings. From this perspective, we can anticipate that MCG\,+04$-$22$-$042, which has three epochs characterized by $R_f \gtrsim 1-3$, would yield a more robust estimate of $\theta_\mathrm{disk}$ than PG\,1310$-$108, which only has an upper limit of $R_f < 2.02$. Additionally, $A_\mathrm{Fe}$ and $kT_e$ of MCG\,+04$-$22$-$042 are also higher than those of PG\,1310$-$108, consistent with the more prominent reflection continuum in the former, even though the effect of these factors is not as significant as the reflection fraction. 

However, if the comparison is drawn between PG\,2209+184 and PG\,1310$-$108, where the former has a slightly bigger error on $\theta_\mathrm{disk}$, the conclusion is less straightforward. Two sources have roughly equal $A_\mathrm{Fe}$. Yet, the coronal temperature of PG\,2209+184 is higher than that of PG\,1310$-$108. Despite having one epoch of simultaneous XMM-Newton and NuSTAR observations compared with the single XMM-Newton epoch of PG\,1310$-$108, PG\,2209+184 has a larger error on $\theta_\mathrm{disk}$. Here, $R_f$ may play a more critical role than data combination in explaining this discrepancy, as PG\,2209+184 has $R_f \approx 0.42$, while PG\,1310$-$108 has $R_f \approx 0.88$ as the 50\% value on the MCMC chain.

\subsubsection{Systematic Uncertainties}

As for systematic concerns, the main problem is the correct attribution of the intrinsic, reflected spectra and other contaminants, for instance, reprocessed emission like absorbers and thermal and nonthermal radiation directly from the accretion flow. We address the bias introduced by absorption in Section~\ref{sec:absorption}, and the bias introduced by soft excess modeling in Section~\ref{sec:soft}. Here, we discuss the potential biases from the reflection model itself.

Theoretically, three factors introduce degeneracies to inclination measurements and the overall spectral fitting. The $\theta_\mathrm{disk}$ in reflection models is degenerate with the spin and the radial emissivity profile due to their mutual effects on producing the broad Fe~K$\alpha$ line and the Compton hump, and likely the soft excess. Atop these, the broadband spectra are tuned by the ionization state of the disk nonmonotonically \citep[e.g., see][]{Bonson and Gallo 2016,Reynolds 2021}, which might lead to equally satisfactory fits with both a high and a low ionization state.

We have found that $a_\ast$ has little effect on the recoverability of $\theta_\mathrm{disk}$ by fitting generated mock spectra with known input parameters \citep[Section~5.4 in][]{Du et al. 2024}, though the constraints on $a_\ast$ itself and the complicated interplay between $a_\ast$ and spectral features are still open topics \citep[e.g.,][]{Reynolds 2021}.

The radial emissivity profile is largely determined after selecting a flavor of coronal configuration within the \textsc{RELXILL} where the profile can be parameterized for various geometries \citep[e.g.,][]{Dauser et al. 2013}. For our choice of \textsc{RELXILLCP}, the empirical emissivity scales at $r^{-\mathrm{Index}_\mathrm{1}}$ between $R_\mathrm{in}$ and $R_\mathrm{br}$ and $r^{-\mathrm{Index}_\mathrm{2}}$ between $R_\mathrm{br}$ and $R_\mathrm{out}$. The usually discovered high inner emissivity index (Index$\mathrm{1} \gtrsim 5$ to 7; as also in our Table~\ref{tab:result_refl}) has motivated the lamppost geometry (as in \textsc{RELXILLLP}) which describes an isotropically irradiating source located on the rotational axis of the BH in its vicinity \citep[e.g.,][]{Wilkins and Fabian 2011,Dauser et al. 2012,Miller et al. 2015,Beuchert et al. 2017}. However, the configuration of the corona is still largely unclear and difficult to distinguish from spectral analysis especially when intending to systematize the methodology on large samples \citep[e.g.,][]{Bonson and Gallo 2016,Tortosa et al. 2018}, even though several attempts in X-ray reverberation mapping \citep[e.g.,][]{Reynolds et al. 1999,Fabian et al. 2009,Kara et al. 2016,Caballero-Garcia et al. 2020} and spectropolarimetry as performed by the Imaging X-ray Polarimetry Explorer (IXPE, \citealp{Weisskopf et al. 2022};  e.g., \citealp{Tagliacozzo et al. 2023,Gianolli et al. 2024a,Gianolli et al. 2024b,Mondal et al. 2024,Serafinelli et al. 2024}, however, see \citealp{Ingram et al. 2023}) have showcast their potential in resolving this intricacy with additional information beyond spectroscopy. For our purpose, adopting \textsc{RELXILLCP} with two emissivity indices and the broken radius implicitly takes into consideration of the coronal geometry without additional assumptions but increases the flexibility of the model than applying the lamppost geometry where several corona-related parameters are directly computed, and thus, careful source by source analysis (as in Appendix~\ref{app:details}) is needed for $\theta_\mathrm{disk}$ measurements based on the reflection model.

For the degeneracy associated with $\xi$, we should also take on careful approaches to explore the parameter space of $\xi$ source by source to reduce possible biases. This problem is also entangled with other absorbing or obscuring components, such as the WAs. But, according to \cite{Kammoun et al. 2018}, if the absorbers are added correctly for high-quality broadband spectra exhibiting notable reflection features (e.g., $a_\ast > 0.8$, and lamppost corona height $h > 5\,r_g$), the spectral fits should not be significantly affected.

\subsubsection{Other Concerns}

Regarding the validity of the reflection model itself, the simulation recipes are based on a thin constant density slab in local thermal equilibrium. However, real accretion disks exhibit more complex structures, including variations in the density profile, vertical stratification, magnetic fields, and radiation pressure, none of which are taken into account in current models. The impact of these factors on the reflection spectrum remains uncertain. Moreover, the accuracy of the reflection model is limited by the quality of the atomic data and radiative transfer calculations \citep[e.g.,][]{Garcia et al. 2013, Ding et al. 2024}. For instance, parameters such as $A_\mathrm{Fe}$, $\theta_\mathrm{disk}$, and disk density might be biased even if the model achieves a statistically acceptable fit. This limitation arises due to the omission of plasma physics effects that become significant in the soft energy regime at accretion disk densities of $\sim 10^{15}-10^{22}\,\rm cm^{-3}$ \citep{Ding et al. 2024}.

Strictly speaking, the $\theta_\mathrm{disk}$ measurements pertain to the innermost regions of the accretion disk, where the gravitational field of the black hole strongly influences the surrounding gas, leading to the broadening of emission lines. These broadened profiles encode the velocity field projection, and thus provide a measurement of the inclination of gas orbits. Therefore, our analysis, which correlates $\theta_\mathrm{BLR}$ with $\theta_\mathrm{disk}$, implicitly examines the underlying relationship between the BLR and the inner accretion disk. It is worth noting that the accretion disks could be tilted or even fragmented into multiple sub-disks \citep[see, e.g., simulations of tilted disks in][]{Liska et al. 2021}. Such structural complexities are not considered in the standard physical models of either the BLR or the X-ray reflection framework, nor are they directly addressed in this study.

\begin{deluxetable}{cc}[t]
    \tablenum{4}
    \caption{Parameter Values of the Simulation Grid\label{tab:sim_grid}}
    \tablehead{
    \colhead{Parameter} &
    \colhead{Input Range}
    }
    \startdata
    $\log N_\mathrm{H}$ & \{19, 20, 21, 22, 23\} \\
    $\log \xi_\mathrm{WA}$ & \{$-2$, $-1$, 0, 1, 2, 3\} \\ 
    $v/c$ & $-0.05$ \\ \hline
    $\theta_\mathrm{disk}$ & \{5, 25, 45, 65, 85\}$^\circ$\\
    $a_\ast$ & \{$\pm0.998$, $\pm0.9$, $\pm0.5$, $0$\} \\
    $R_f$ & \{0, 0.3, 1, 5, 10\} \\
    $A_\mathrm{Fe}$ & 3\,$A_\odot$\\
    $\Gamma$ & 2\\
    $\log \xi$ & 3\\
    $kT_{e}$ & 85\,keV\\
    Index$_\mathrm{1}$ & 7\\
    Index$_\mathrm{2}$ & 3\\
    $R_\mathrm{br}$ & 10\,$r_{g}$
    \enddata
\end{deluxetable}

\begin{figure*}[t]
    \centering
    \includegraphics[width=0.9\linewidth]{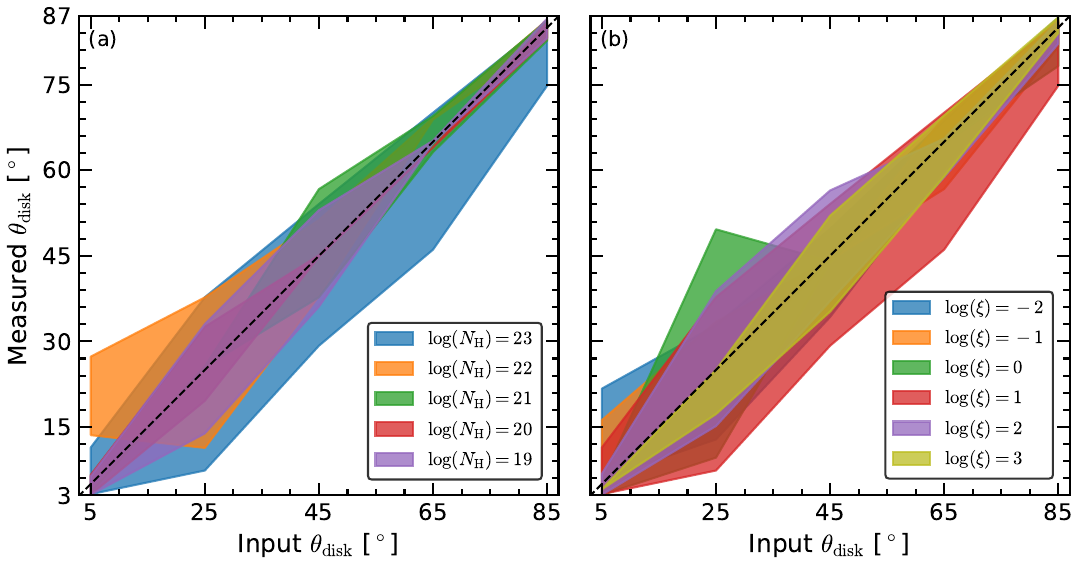}
    \caption{Inner accretion disk inclination measurements for simulated spectra with different WA values of (a) column density and fixed WA ionization at $\log \xi_\mathrm{WA} = 1$, and (b) different values of WA ionization $\xi_\mathrm{WA}$ and fixed column density at $\log N_\mathrm{H} = 23$. In all panels, we are showing the slice of parameter space with spin $a_\ast=0.998$ and reflection fraction $R_f=1$.}
    \label{fig:incl_wa}
    \end{figure*}

\subsection{Bias from the Warm Absorbers} \label{sec:absorption}

Modeled with \textsc{XSTAR} tables, the WAs in our spectral fittings reprocess the intrinsic spectra, potentially mimicking the relativistic effects that influence $\theta_\mathrm{disk}$. To assess the impact of WAs on inclination measurements, we perform analyses on mock spectra simulating a single-epoch simultaneous observation of a nearby type~1 AGN. This approach evaluates the systematic biases introduced by the WAs in parameter retrievals. Alternatively, X-ray reverberation mapping could be employed to disentangle the reflection spectrum from absorption features \citep[e.g.,][]{Fabian et al. 2009, Kara et al. 2016}, but it is beyond the scope of this work.

Similar to the analysis in Section~4 of \cite{Du et al. 2024}, we generate mock spectra with the baseline model \textsc{CONSTANT*XSTAR*(ZBBODY+RELXILLCP)} using the \texttt{fakeit} method in \textsc{PyXspec} \citep{Arnaud 1996,Arnaud 2016}. The mock spectra are produced in the 0.3--10\,keV band of XMM-Newton/EPIC and the 3--78\,keV band of NuSTAR, with background, response, and ancillary files from an observation epoch of 3C\,382 (XMM-Newton: 0790600301, NuSTAR: 60202015006). The source exposure and background exposure times are set to 25\,ks. The inner radius of the accretion disk is set to the radius of the innermost stable circular orbit; the outer radius is set to $R_\mathrm{out} = 400\,r_{g}$; and the density is set to $n = 10^{15}\,\mathrm{cm^{-3}}$. The blackbody temperature is set to 0.1\,keV. The redshift is set to $z=0.05557$ \citep{van den Bosch et al. 2015}. Other parameters are initialized as shown in Table~\ref{tab:sim_grid}, from which we create 5,250 distinct spectra.

The column density ($N_\mathrm{H}$) and ionization parameter ($\xi_\mathrm{WA}$) of the WA span the physical ranges typically observed in AGNs \citep[e.g.,][]{Laha et al. 2021}. The velocity is set to $v/c = -0.05$, consistent with commonly observed outflowing features \citep[e.g.,][]{Braito et al. 2014}. The inclination angle is varied from nearly face-on to nearly edge-on configurations. BH spin ranges from $+0.998$ to $-0.998$, covering extreme prograde to non-rotating (Schwarzschild), and to extreme retrograde cases. Reflection fractions are set to 0, 0.3, 1, 5, and 10, representing non-reflection, ``reflection-subordinate'', equal reflection and continuum, ``reflection-dominant'', and highly reflected scenarios, respectively \citep{Du et al. 2024}. The iron abundance is set to three times the solar value, with a photon index of two and an ionization parameter of three, all consistent with observed AGN ranges. The coronal temperature $kT_{e}$ is fixed at 85\,keV, adopted around the average of measurements in previous study \citep{Kamraj et al. 2022}. The emissivity profile is modeled as a broken power law with indices of  Index$_\mathrm{1} = 7$ and Index$_\mathrm{2} = 3$, and the broken radius $R_\mathrm{br} = 10\,r_{g}$.

The mock spectra are fit using the same baseline model with technical setups applied for single-epch, joint XMM-Newton and NuSTAR observations. Given the known input parameters, initiating the fit directly at the input values or randomizing the initial conditions for the MCMC error calculations is inappropriate \citep[e.g.,][]{Bonson and Gallo 2016, Choudhury et al. 2017}. Instead, we perform a preliminary fit and configure the proposal distribution for the MCMC chains in \textsc{XSPEC} as a Gaussian with a diagonal covariance matrix derived from 100 times the step length of the parameters in the preliminary fit. This approach ensures the sampler to explore the parameter space but without becoming trapped in local minima, especially for the flexible and complex baseline model in use. The MCMC chains are run with 200 walkers, generating at least $6 \times 10^5$ elements, with the first 10\% discarded as burn-in to ensure convergence.

Building on the discussion in \cite{Du et al. 2024} regarding the influence of $R_f$, $A_\mathrm{Fe}$, $kT_e$, and $a_\ast$ on $\theta_\mathrm{disk}$ measurements, our analysis focuses specifically on the effects of WAs. Results from two slices of the simulation grid are visualized in Figure~\ref{fig:incl_wa} as representative examples. Overall, the inclination measurements are robust against the inclusion of WAs. The column density of the WA exhibits a minor effect, with errors in $\theta_\mathrm{disk}$ decreasing as $N_\mathrm{H}$ increases. This trend, which worsens measurements for denser WAs, reflects competition between intrinsic reflection and reprocessed emission in the spectral fitting. However, this effect becomes significant only at extreme column densities, $\log N_\mathrm{H} = 22$ to 23, and primarily affects only $\theta_\mathrm{disk}$, with negligible impacts on $a_\ast$ and $R_f$. The WA ionization parameter has an insignificant influence on the results. Across the entire simulation grid of 5,250 spectra, 94\% exhibit an absolute inclination offset smaller than $5^\circ$ between measured and input values; 98\% have offsets smaller than $10^\circ$; and 54\% have 90\% confidence intervals that encompass the input value with offsets smaller than $5^\circ$. All reduced fit statistics are below 1.05.

The most robust approach to studying parameter behavior is to generate mock spectra with identical spectral coverage to actual observations and with known input parameters over a fine grid encompassing the parameter space of interest, followed by blind fitting with the same model. However, this method is computationally intensive and time-consuming, necessitating practical compromises. Previous studies \citep{Bonson and Gallo 2016, Choudhury et al. 2017, Kammoun et al. 2018, Du et al. 2024} have systematically tested parameter recovery with mock spectra. While the first three focused on $a_\ast$ accuracy, \cite{Du et al. 2024} emphasized $\theta_\mathrm{disk}$ reliability. Notably, only \cite{Kammoun et al. 2018} incorporated WA effects, using 60 mock spectra with more complex components, including warm and neutral absorbers, relativistic (\textsc{RELXILLLP}) and distant (\textsc{XILLVER}) reflection, and thermal emission. Their simulations involved three collaborators: each of the one generating the spectra and two others independently fitting them, closely mimicking real observational procedures. They concluded that neither absorbers nor $a_\ast$ significantly influenced the fits, though the lamppost height in \textsc{RELXILLLP} did. Our work builds on these efforts, employing a larger simulation grid with a simplified fitting and error calculation approach compared to \cite{Kammoun et al. 2018}. However, our primary aim is to evaluate the reliability of $\theta_\mathrm{disk}$ measurements rather than exhaustively exploring the entire parameter space or probing the recovery and degeneracies of all baseline model parameters.

\begin{deluxetable*}{l|cccc|cccccc}
    \tablenum{5}
    \caption{Parameters Used to Interpret the Soft Excess\label{tab:soft_excess}}
    \tablehead{
    \colhead{Name} &
    \colhead{$kT$} &
    \colhead{$\theta_\mathrm{disk}$} &
    \colhead{$\chi^2$} &
    \colhead{DOF} &
    \colhead{$\Gamma_\mathrm{c}$} &
    \colhead{$kT_\mathrm{ce}$} &
    \colhead{$kT_\mathrm{cbb}$} &
    \colhead{$\theta_\mathrm{disk}$} &
    \colhead{$\chi^2$} &
    \colhead{DOF} \\
    & \colhead{(keV)} & 
    \colhead{($^\circ$)} & & & &
    \colhead{(keV)} & 
    \colhead{(eV)} &
    \colhead{($^\circ$)} & &\\
    \colhead{(1)} &
    \colhead{(2)} &
    \colhead{(3)} &
    \colhead{(4)} &
    \colhead{(5)} &
    \colhead{(6)} &
    \colhead{(7)} &
    \colhead{(8)} &
    \colhead{(9)} &
    \colhead{(10)} &
    \colhead{(11)} 
    }
    \startdata
    MCG\,+04$-$22$-$042 & $0.165_{-0.022}^{+0.034}$ & $15.2_{-1.5}^{+2.2}$ & 1651.6 & 1655 & $9.37_{-0.34}^{+0.31}$ & $1.62_{-0.09}^{+0.08}$ & $224_{-11}^{+10}$ & $16.5_{-0.3}^{+0.2}$ & 1650.4 & 1653 \\
    Mrk\,279 & $0.089 \pm 0.004$ & $26.9_{-3.7}^{+2.5}$ & 1327.9 & 1186 & $8.50_{-0.10}^{+0.09}$ & $1.62_{-0.08}^{+0.09}$ & $106_{-2}^{+4}$ & $25.8_{-1.0}^{+0.9}$ & 1337.3 & 1184 \\
    Mrk\,1392 & $0.040_{-0.008}^{+0.005}$ & $25.5_{-9.9}^{+6.7}$ & 453.1 & 401 & $8.65 \pm 0.13$ & $42.5_{-0.4}^{+0.5}$ & $22.4_{-0.2}^{+0.1}$ & $26.2_{-0.5}^{+0.1}$ & 454.1 & 399 \\
    PG\,1310$-$108 & $0.089_{-0.010}^{+0.007}$ & $33.5_{-22.4}^{+20.8}$ & 389.6 & 373 & $4.13 \pm 0.21$ & $2.13_{-1.10}^{+8.08}$ & $1.4_{-0.4}^{+8.8}$ & $38.7_{-2.1}^{+0.8}$ & 435.1 & 371 \\
    PG\,2209+184 & $0.086_{-0.017}^{+0.015}$ & $35.6_{-25.9}^{+31.7}$ & 1023.0 & 915 & $4.63_{-0.64}^{+0.86}$ & $9.62_{-8.29}^{+28.9}$ & $7.8_{-6.7}^{+46.0}$ & $40.3_{-8.9}^{+6.1}$ & 1031.5 & 913 \\
    RBS\,1917 & $0.248_{-0.031}^{+0.009}$ & $17.2_{-1.1}^{+4.5}$ & 410.7 & 351 & $2.50 \pm 0.01$ & $> 300$ & $139 \pm 1$ & $17.0 \pm 0.01$ & 490.1 & 349 \\
    Zw\,229$-$015 & $0.103 \pm 0.001$ & $26.2_{-0.5}^{+2.5}$ & 471.2 & 463 & $3.77_{-0.02}^{+0.04}$ & $1.62 \pm 0.01$ & $5.4 \pm 0.1$ & $29.0_{-0.5}^{+0.8}$ & 479.4 & 461
    \enddata
    \tablecomments{
    Col. (1): Name of the source.
    Col. (2): Temperature of the epoch-independent blackbody (see Appendix~\ref{app:details}).
    Col. (3): Inclination of the inner accretion disk when the blackbody is used (Column~3 of Table~\ref{tab:result_refl}).
    Col. (4): $\chi^2$ statistic of the fit with the blackbody.
    Col. (5): Degrees of freedom of the fit with the blackbody.
    Col. (6): Power-law photon index of the Comptonization component (\textsc{NTHCOMP}).
    Col. (7): Electron temperature of the Comptonization component.
    Col. (8): Seed photon temperature of the Comptonization component.
    Col. (9): Inclination of the inner accretion disk when the Comptonization component is used.
    Col. (10): $\chi^2$ statistic of the fit with the Comptonization component.
    Col. (11): Degrees of freedom of the fit with the Comptonization component.
    }
\end{deluxetable*}

\subsection{Bias from Treatments of the Soft Band} \label{sec:soft}

In our spectral fits, an epoch-independent blackbody is employed to model a portion of the soft excess. While effective, this is a simplified approach, as the exact nature of the soft excess remains a topic of active debate \citep[e.g.,][]{Crummy et al. 2006, Done et al. 2012, Jin et al. 2017, Garcia et al. 2019, Gliozzi and Williams 2020, Petrucci et al. 2020}. Current explanations for the soft excess include relativistically smeared reflection features at soft energies \citep[e.g.,][]{Garcia et al. 2019}, Comptonization in a warm corona \citep[e.g.,][]{Petrucci et al. 2020}, or a combination of both. Additionally, complex absorption and scattering phenomena in the soft band can mimic relativistic effects typically associated with reflection, as discussed in Section~\ref{sec:absorption}.

In our analysis, the soft excess has been modeled with a simple, phenomenological redshifted blackbody. This approach has been demonstrated to work effectively in Section~5.7 of \cite{Du et al. 2024}, provided that the WAs are accurately characterized using high-quality broadband spectra. In this section, we evaluate how different treatments of the soft excess influence inclination measurements.

Since Mrk\,50 is well-represented by pure reflection without requiring a blackbody or WA component, we exclude it from this test. For all other sources, we refit the spectra using the Comptonization component \textsc{NTHCOMP} \citep{Zdziarski et al. 1996, Zycki et al. 1999} in \textsc{XSPEC} to model the soft excess in place of \textsc{ZBBODY}. The asymptotic power-law photon index $\Gamma_\mathrm{c}$, electron temperature $kT_\mathrm{ce}$, and seed photon temperature $kT_\mathrm{cbb}$ are set as free parameters, with the seed photons assumed to originate from a disk blackbody. The reflection model remains unchanged, meaning that the substitution introduces two additional parameters into the model. Errors are computed using the MCMC approach described in Section~\ref{sec:fitting}.

Table~\ref{tab:soft_excess} summarizes the results, including the blackbody temperature ($kT$), $\theta_\mathrm{disk}$, and $\chi^2$ values for the fits. Substituting the blackbody with the Comptonization component does not lead to significant deviations in the inclination measurements. Similarly, $R_f$ and $\xi$ remain consistent with the baseline model. Despite the addition of two free parameters, the fit statistics are not substantially improved—and in some cases, slightly worsen—indicating potential overfitting or insufficient physical motivation for the Comptonization component.

In conclusion, based on current data quality, the treatment of epoch-independent blackbody provides a sufficiently accurate representation of the soft excess, with negligible influence on inclination measurements compared to the Comptonization component at the current level of spectral quality and fit precision. Nevertheless, we emphasize the importance of accurate soft-band modeling in enhancing the precision of reflection model fits. Arbitrary treatments of the soft band should be avoided, as they could inadvertently affect the reliability of $\theta_\mathrm{disk}$ measurements.

\begin{figure}
    \centering
    \includegraphics[width=\linewidth]{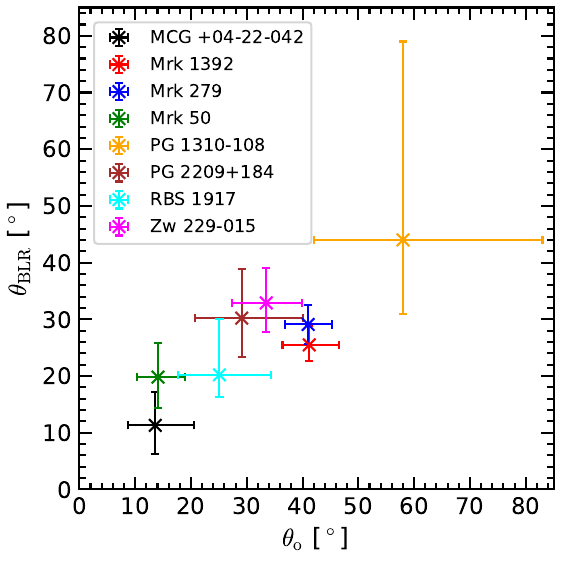}
    \caption{The relation between the BLR inclination versus its opening angle.}
    \label{fig:additional_angles1}
\end{figure}

\begin{figure}
    \centering
    \includegraphics[width=\linewidth]{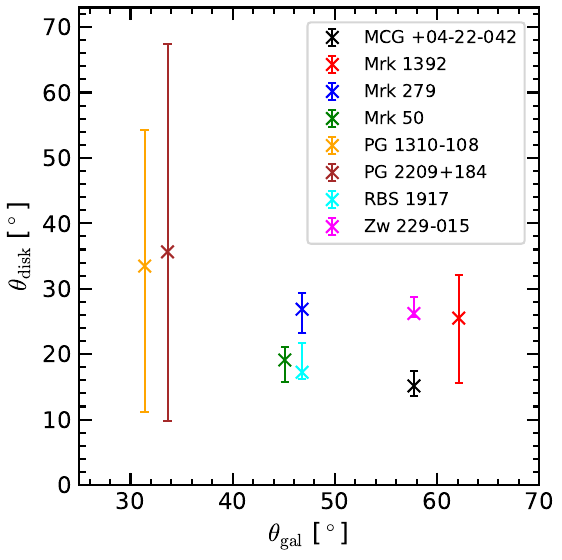}
    \caption{The relation between the inclination of the inner accretion disk versus the large-scale inclination of the galactic disk. Errors for the galactic disk inclination are not shown.}
    \label{fig:additional_angles2}
\end{figure}

\subsection{Error on the BLR Inclination}

Similar to the accretion disk measurements, the errors of the BLR inclination angles also vary among the sources. Under most circumstances, the \textsc{CARAMEL} models give formal errors of $5^\circ-10^\circ$ for $\theta_\mathrm{BLR}$ (Table~\ref{tab:angles}). However, the full dynamical model of the BLR is complex and has many parameters and underlying assumptions whose uncertainties are difficult to quantify. For instance, the \textsc{CARAMEL} framework ignores the effect of radiation pressure on the dynamics of the gas. As the radiation pressure has a radial dependence similar to gravity, omitting it can underestimate the BH mass \citep{Marconi et al. 2008,Pancoast et al. 2014a, Villafana et al. 2022}. The model adopts simplified dynamics for point particles, neglecting detailed structures of the BLR and other lower order factors such as self-gravity, gas viscosity, and interactions between the BLR and accretion disk gas. The actual BLR might have multiple sub-structures with different geometries and dynamics \citep[e.g.,][]{Baldwin et al. 1995,Collin et al. 2006}, each with slightly different inclination and kinematics. By necessity, the inclination results we use are only phenomenological averages of the BLR traced by the H$\beta$ line. This oversimplification obviously can introduce subtle errors in $\theta_\mathrm{BLR}$.

Another complication is that the other geometric parameter in the model---the opening angle $\theta_\mathrm{o}$---can be related closely to and hence affecting the precision of $\theta_\mathrm{BLR}$. Figure~\ref{fig:additional_angles1} plots $\theta_\mathrm{BLR}$ against $\theta_\mathrm{o}$, where we have taken the opening angle measurements (Table~\ref{tab:angles}) directly from \cite{Williams et al. 2018} and \cite{Villafana et al. 2022}. The Pearson correlation coefficient is 0.882 with a $p$-value of 0.004, indicating a strong positive correlation between $\theta_\mathrm{o}$ and $\theta_\mathrm{BLR}$ at the level of $2.9\,\sigma$. It is thus likely that $\theta_\mathrm{o}$ and $\theta_\mathrm{BLR}$ are not independent in the geometrical description of the system. Naturally, $\theta_\mathrm{BLR}$ is more difficult to constrain precisely when $\theta_\mathrm{o}$ is large. It is hard to distinguish between a geometrically thick BLR that is face-on or edge-on. Furthermore, since the two parameters affect the projection of the velocity field onto the line-of-sight in opposite ways, dynamical modeling of a thick, extended BLR may require a corresponding inclination for compensation in order to match the observed velocity profiles. This correlation was first identified by \cite{Grier et al. 2017}, who noted that the model required a positive relationship between the inclination angle and the opening angle to reproduce the observed emission-line profiles. This dependency arises from the distribution of the transfer function in wavelength and time-lag space. A potential method to disentangle this degeneracy involves interferometric observations, as demonstrated by \cite{Gravity Collaboration et al. 2020}, which could provide more direct constraints on the BLR geometry and kinematics.

There is no significant correlation between the error of $\theta_\mathrm{BLR}$ and the value of $\theta_\mathrm{o}$ (correlation coefficient 0.600, $p$-value 0.116). The moderate correlation suggests that $\theta_\mathrm{o}$ might affect the precision of $\theta_\mathrm{BLR}$ secondarily or non-linearly, as other parameters such as the BLR radial profile and velocity field could also have simultaneous impact, as do purely geometric factors. The correlation coefficient of the error of $\theta_\mathrm{BLR}$ and the error of $\theta_\mathrm{o}$ is 0.977 with a $p$-value of $3\times 10^{-5}$, which is significant at $4.2\,\sigma$. The mutual growth in the size of the error bars is evident in the plot. The very strong correlation between the errors implies that the uncertainties in $\theta_\mathrm{o}$ may have propagated directly into the uncertainties in the inclination angle, and vice versa, suggesting that the BLR geometry is more likely determined by the mutual effects of the inclination and the opening angle on the velocity field projection.

\subsection{Connection to Galactic Scales} \label{sec:galactic}

If the inclination correlation on the scale of the BLR and accretion disk can be confirmed, then inclination measurements on small scales can be used to test for (mis)alignment with the galactic disk on large scales, as a means of probing the fueling mechanism and cosmological evolution of active galaxies \citep[e.g.,][]{Hopkins et al. 2012}. As a demonstration based on this work, we compare the inner accretion disk inclination and BLR inclination with the galactic disk inclination $\theta_\mathrm{gal}$ to examine whether the angular momentum of different scales in active galaxies are related. Here, we assume that the galactic disk inclination serves as a proxy for the large-scale angular momentum of the host galaxy ($\mathbf{J}_\mathrm{gal}$), and that the inclination of the inner accretion disk tracks the small-scale angular momentum of the central BH ($\mathbf{J}_\mathrm{BH}$). The alignment ($\mathbf{J}_\mathrm{gal}   \parallel \mathbf{J}_\mathrm{BH}$ or $\mathbf{J}_\mathrm{gal} \cdot \mathbf{J}_\mathrm{BH} \approx |\mathbf{J}_\mathrm{gal}||\mathbf{J}_\mathrm{BH}|$) or misalignment ($\mathbf{J}_\mathrm{gal} \cdot \mathbf{J}_\mathrm{BH} < |\mathbf{J}_\mathrm{gal}||\mathbf{J}_\mathrm{BH}|$) of the two angular momenta are then simplified to the alignment or misalignment of the two inclinations, which may offer insights into the feeding mechanism of the supermassive BH. Alignment requires systematic, long-term co-planar fueling \citep[e.g.,][]{King et al. 2005}, whereas misalignment might result from mergers that knock the BH spin off the host galaxy plane \citep[e.g.,][]{Gerosa et al. 2015}, if the \cite{Bardeen and Petterson 1975} effect has not yet aligned the accretion flow with the BH spin.

We calculate galaxy inclination angles (Table~\ref{tab:angles}) using $K_s$ band images from the Two Micron All Sky Survey \citep[2MASS,][]{Skrutskie et al. 2006}. The galactic inclination follows from \citep{Hubble 1926} 

\begin{equation}
\cos \theta_\mathrm{gal} = \sqrt{\frac{q^2 - q_0^2}{1 - q_0^2}},
\end{equation}

\noindent
with $q = b/a$ the minor-to-major axis ratio \citep{Jarrett et al. 2000} measured with isophotal photometry set at 20\,mag\,arcsec$^{-2}$ in the $K_s$ band, and $q_0$ the intrinsic axis ratio assumed to be 0.2 \citep[][see \citealt{Yu2020} for a more complex treatment]{Holmberg 1946,Guthrie 1992}. Plotting $\theta_\mathrm{disk}$ against $\theta_\mathrm{gal}$ reveals no obvious trend (Figure~\ref{fig:additional_angles2}). The Pearson correlation coefficient between $\theta_\mathrm{disk}$ and $\theta_\mathrm{gal}$ is $-0.597$ with a $p$-value of 0.118, and the correlation coefficient between $\theta_\mathrm{BLR}$ and $\theta_\mathrm{gal}$ is $-0.540$ with a $p$-value of 0.167. The lack of statistical significance suggests that no meaningful relationship holds between the inclinations on very large and small scales in active galaxies, at least within our limited sample. \citet[][their Section~3.3.3]{Kormendy and Ho 2013} and \cite{Wu et al. 2022} reached a similar conclusion. These indicate the possible existence of a diverse AGN feeding mechanisms of both coplanar accretion and minor mergers.

A closely related investigation that arrived at somewhat different results was conducted by \cite{Middleton et al. 2016}, who found a loose ($3\,\sigma$) correlation between the inner accretion disk inclination based on previous reflection modeling and the galactic stellar disk inclinations mainly taken from \textsc{HyperLEDA} \citep{Makarov et al. 2014} for their sub-sample of 21 AGNs. Yet, their complete sample of 26 AGNs returned lack of correlation at $\gg 5\sigma$. In a recent extensive study of the position angle differences between radio and optical images of radio AGNs, \cite{Zheng2024} conclude that jet alignment depends on radio power and galaxy shape: the jets of low radio power tend to show preferential alignment with the minor axis of optically more flattened host galaxies.

\section{Summary} \label{sec:summary}

We investigate the relationship between the inclination angle of the inner accretion disk and the BLR in AGNs. Starting with type~1 AGNs with published BLR inclination measurements derived from velocity-resolved RM campaigns, we closely follow the methodology of \cite{Du et al. 2024} to analyze the broadband 0.3--78\,keV spectra of a small sample of eight sources with sufficient data for X-ray reflection spectroscopic analysis using the self-consistent reflection model \textsc{RELXILLCP}. The RM results provide measurements of $\theta_\mathrm{BLR}$, while our spectral fits derive $\theta_\mathrm{disk}$. While the two inclinations show a strong, positive correlation (Pearson correlation coefficient 0.856, $p$-value 0.007), the correlation is marginal ($<3\,\sigma$) according to our Monte Carlo simulations. Nevertheless, a nearly linear relation between $\theta_\mathrm{BLR}$ and $\theta_\mathrm{disk}$ revealed by our regression analysis hints at a possible alignment between the accretion disk and the BLR, which has a number of implications, including the possibility of elucidating physical models that link these two fundamental components of the central engine of active galaxies. An alignment between these two fundamental components of AGNs would suggest that the rotational axes of the BLR and the accretion disk are likely the same. Moreover, on account of the Bardeen-Petterson effect, this common rotation axis should also be the orientation with respect to the line-of-sight for the BH spin. If more sources with $\theta_\mathrm{disk}$ spanning $0-\pi/2$ can be added in the future to support the correlation, we would be able to claim that the BLR, believed to emanate from near or above the accretion disk, is closely related to the accretion disk or potentially shaped by the same axisymmetric environment as the accretion disk.

While disk wind models are often invoked to explain the formation of the BLR \citep[e.g.,][]{Peterson 2006}, debate continues as to whether the disk winds derive from radiation pressure-driven outflows or magnetically driven outflows \citep[e.g., see][]{Czerny and Hryniewicz 2011}. In any case, the disk winds should connect the BLR with the accretion disk in many physical aspects. We note, in passing, that all seven of the sources in our sample that show evidence for WAs (Table~\ref{tab:absorbers}) have outflows, whose velocities exceed $0.1\,c$ in four cases (MCG\,+04$-$22$-$042, Mrk\,1392, PG\,1310$-$108, and Zw\,229$-$015). This could be expected if BLR clouds indeed form from launched disk winds, similar to those that produce the WAs. Previous BLR dynamical modeling of RM observations \citep{Pancoast et al. 2014b} find that the broad emission lines on the far side of the accretion disk are hidden by optically thick material, which can plausibly be associated with disk winds \citep[e.g.,][]{Shields 1977,Murray and Chiang 1995,Elvis 2017} because WAs are observed as blueshifted absorption lines in both the X-ray and UV.

\section*{Acknowledgements}
We thank the anonymous referee for helpful suggestions. This work was supported by the National Key R\&D Program of China (2022YFF0503401), the National Science Foundation of China (11991052, 12233001), and the China Manned Space Project (CMS-CSST-2021-A04, CMS-CSST-2021-A06). This research has made use of the NASA/IPAC Extragalactic Database (NED), which is funded by the National Aeronautics and Space Administration and operated by the California Institute of Technology. We thank Jinyi Shangguan for helpful comments.

\vspace{5mm}
\facilities{XMM-Newton, NuSTAR}

\software{\textsc{AstroPy} \citep{Astropy Collaboration et al. 2013,Astropy Collaboration et al. 2018,Astropy Collaboration et al. 2022}, \textsc{EMCEE} \citep{Foreman-Mackey et al. 2013}, \textsc{Matplotlib} \citep{Hunter 2007}, \textsc{Numpy} \citep{van der Walt et al. 2011, Harris et al. 2020}, \textsc{SciencePlots} \citep{Garrett 2021}, \textsc{XSPEC} \& \textsc{PyXspec} \citep{Arnaud 1996,Arnaud 2016}
}

\appendix

\section{Details on X-Ray Spectroscopy} \label{app:details}

In this appendix, we provide the details of our spectral fitting of the eight sources described in Sections~\ref{sec:sample} and \ref{sec:data_reduction}, where the general procedure follows Section~\ref{sec:fitting}. We summarize the best-fit values of the continuum in Table~\ref{tab:result_refl} and the results of WAs in Table~\ref{tab:absorbers}.

\begin{deluxetable}{crrc}
    \tablenum{A1}
    \tablecaption{Best-fit Parameters for the Warm Absorbers}
    \tablehead{
    \colhead{Source} &
    \colhead{$N_\mathrm{H}$} &
    \colhead{$\log \xi_\mathrm{WA}$} &
    \colhead{$v/c$} \\
    & \colhead{($10^{20}\,\rm cm^{-2}$)} & \colhead{($\mathrm{erg~cm~s^{-1}}$)} & \\
    \colhead{(1)} & \colhead{(2)} & \colhead{(3)} & \colhead{(4)}
    }
    \startdata
    MCG\,+04$-$22$-$042 & $12.92_{-3.32}^{+3.30}$ & $1.42_{-0.20}^{+0.23}$ & $0.114_{-0.015}^{+0.011}$ \\ \hline
    Mrk\,279 & $30.46_{-4.51}^{+6.65}$ & $2.44_{-0.10}^{+0.11}$ & $0.057 \pm 0.009$ \\ \hline
    Mrk\,1392 & $7.83_{-1.90}^{+2.87}$ & $-1.52_{-0.17}^{+0.26}$ & $0.129_{-0.045}^{+0.040}$ \\ \hline
    PG\,1310$-$108 & $3.12_{-1.71}^{+6.74}$ & $0.33_{-0.77}^{+1.12}$ & $0.181_{-0.025}^{+0.029}$ \\ \hline
    PG\,2209+184 & $3.87_{-2.67}^{+4.66}$ & $1.56_{-0.48}^{+0.66}$ & $0.041_{-0.034}^{+0.032}$ \\ \hline
    RBS\,1917 & $3.96_{-3.28}^{+7.03}$ & $0.70_{-0.41}^{+0.43}$ & $0.081_{-0.016}^{+0.004}$ \\ \hline
    Zw\,229$-$015 & $2.92_{-0.24}^{+0.07}$ & $0.22_{-0.05}^{+0.01}$ & $0.163_{-0.006}^{+0.001}$
    \enddata
    \tablecomments{
        Col. (1): Name of the source.
        Col. (2): Hydrogen column density.
        Col. (3): Ionization parameter.
        Col. (4): Outflow velocity divided by the speed of light.
        The best-fit values and 90\% confidence intervals are presented.}
    \label{tab:absorbers}
\end{deluxetable}

\begin{figure}[t]
    \centering
    \includegraphics[width=1.00\linewidth]{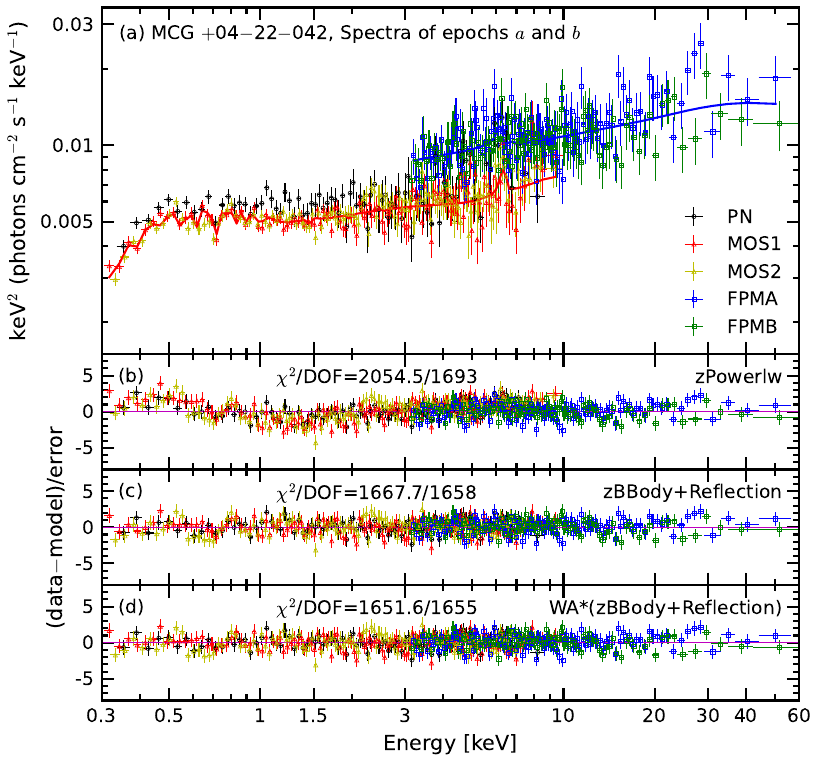}
    \caption{The XMM-Newton and NuSTAR broadband spectral fit of MCG\,+04$-$22$-$042, showing the best fit with the baseline model (a). All epochs are fit simultaneously. The residuals in terms of error sigmas are plotted for (b) the phenomenological redshifted power-law model, (c) \textsc{RELXILLCP} plus a redshifted blackbody component, and (d) \textsc{RELXILLCP} plus a redshifted blackbody component and a WA. For clarity, only the spectra of epoch $a$ (PN, MOS1, MOS2) and epoch $b$ (FPMA, FPMB) are shown.}
    \label{fig:mcgp04m22}
\end{figure}

\subsection{MCG\,+04$-$22$-$042}
MCG\,+04$-$22$-$042 was observed by XMM-Newton once in 2006 for 122\,ks (0312191401) but less than 10\% of the observation time was effective. It was also observed by NuSTAR in 2012 (60061092002), 2020 (60602018002), March 2021 (60602018004), and June 2021 (60602018006). We use all five observation epochs ($a-e$). The NuSTAR data, even though not collected simultaneously with the XMM-Newton observation, provides information on the Compton hump and improves the fit of the entire reflection continuum. During the fit, we tie the disk-related parameters ($\theta_\mathrm{disk}$, $a_\ast$, and $A_\mathrm{Fe}$) and fit the data jointly while setting free the corona-related parameters for each epoch.

We illustrate the broadband fitting procedure in Figure~\ref{fig:mcgp04m22}. In terms of reduced statistic (fit statistic over degrees of freedom, DOF), the blackbody plus reflection continuum improves the fit by $\Delta \chi^2/\Delta\mathrm{DOF} = -386.8/-35$ compared to the phenomenological power-law \textsc{ZPOWERLW}. Further adding the WA returns $\chi^2/\mathrm{DOF} = 1651.6/1655 = 0.998 \:(\Delta \chi^2/\Delta\mathrm{DOF} = -16.1/-3)$, which gives the best fit for this source. Although the likely diverse nature of the soft excess cannot be explained uniquely by the reflection model \citep{Du et al. 2024}, we suggest that an epoch-independent blackbody (\textsc{ZBBODY}), together with \textsc{RELXILLCP}, offers an acceptable description of the soft excess in most situations. In the case of MCG\,+04$-$22$-$042, the X-ray spectrum can be fit by a \textsc{ZBBODY} model with the blackbody temperature of $0.165_{-0.022}^{+0.034}$\,keV. Subtracting this component from the best-fit model returns an inferior statistic of $\Delta \chi^2/\Delta\mathrm{DOF} = +10.1/+2$. Our reflection model results in $\theta_\mathrm{disk} = 15^\circ.2_{-1.5}^{+2.2}$ and $a_\ast = 0.996_{-0.013}^{+0.002}$. Because the refection fractions of all five epochs are not very small \citep[$0.3 \lesssim R_f \lesssim 3$; see Section~5.1. of][where $R_f < 0.3$ is considered reflection-subordinate]{Du et al. 2024} and all five epochs are fit jointly, the model values are calculated with rather high precision.

Three studies on X-ray properties of MCG\,+04$-$22$-$042 used XMM-Newton or NuSTAR. \cite{Winter et al. 2008} analyzed our first spectral epoch~($a$) and suggested that this XMM-Newton observation, together with another non-simultaneous epoch from the Swift X-Ray Telescope \citep[XRT;][]{Gehrels et al. 2004,Burrows et al. 2005}, were well interpreted using an absorbed power-law model with pegged normalization (\textsc{PEGPWRLW}) at the photon index of $\Gamma \approx 2.00$. However, with their reduced statistic of $\chi^2/\mathrm{DOF} = 1534.8/1190 = 1.290$, their fit is not strongly favored statistically. \cite{Akylas and Georgantopoulos 2021} analyzed the NuSTAR spectra that belongs to our epoch~$b$ using a reflection model \textsc{PEXMON} \citep{Nandra et al. 2007}, which describes the neutral reflection of an exponentially cutoff power-law spectrum while self-consistently generating narrow Fe~K lines. Fixing the disk inclination to $60^\circ$, they obtained a reduced statistic of 0.93 with $\Gamma = 1.92 \pm 0.06$ and a reflection scaling factor (equivalent to $R_f$) of $R = 0.53_{-0.22}^{+0.24}$. \cite{Kang and Wang 2022} interpreted the same NuSTAR data set with \textsc{PEXRAV} \citep{Magdziarz and Zdziarski 1995}, which also accounts for a neutral reflector. Assuming solar abundances and $\cos \theta_\mathrm{disk} = 0.45$, they derived $\Gamma = 1.95_{-0.09}^{+0.10}$ and $R = 0.59_{-0.33}^{+0.44}$ with $\chi^2/\mathrm{DOF} = 0.87$. Additionally, they used \textsc{RELXILLCP} to calculate the coronal temperature of $kT_e > 37\, \rm keV$ at $\chi^2/\mathrm{DOF} = 0.88$ while adopting maximal Kerr spin, neutral ionization state, and a disk inclination of $30^\circ$. Since their fits returned small reduced statistics, they may have underestimated the uncertainties of the parameters. Besides, notice that they arrived at a smaller lower limit of the coronal temperature than ours when using the same model but assuming the small $\xi = 0$ and the $\theta_\mathrm{disk}$ around twice as large as ours. As another recent reflection spectroscopic analysis yet using data aside from the two observatories, \cite{Waddell and Gallo 2020} studied data from the Suzaku satellite \citep{Mitsuda et al. 2007} and modeled the continuum with a power-law, a Compton hump, an Fe~K$\alpha$ line, and a soft excess represented by a blackbody. They obtained a photon index of $1.88 \pm 0.02$ and a blackbody temperature ($kT = 0.13 \pm 0.01$\,keV) similar to ours, though with a slightly disfavored reduced \cite{Cash 1979} statistic of $C$-stat$/\mathrm{DOF} = 290/248 = 1.17$. Despite different treatments of the reflection continuum, the results from these works are generally consistent with our findings. However, our analysis is conducted with as many as five epochs while setting as many parameters free as possible, including $a_\ast$ and $\theta_\mathrm{disk}$ which result in $0.996_{-0.013}^{+0.002}$ and $15^\circ.2_{-1.5}^{+2.2}$, respectively.

\begin{figure}[t]
\centering
\includegraphics[width=1.00\linewidth]{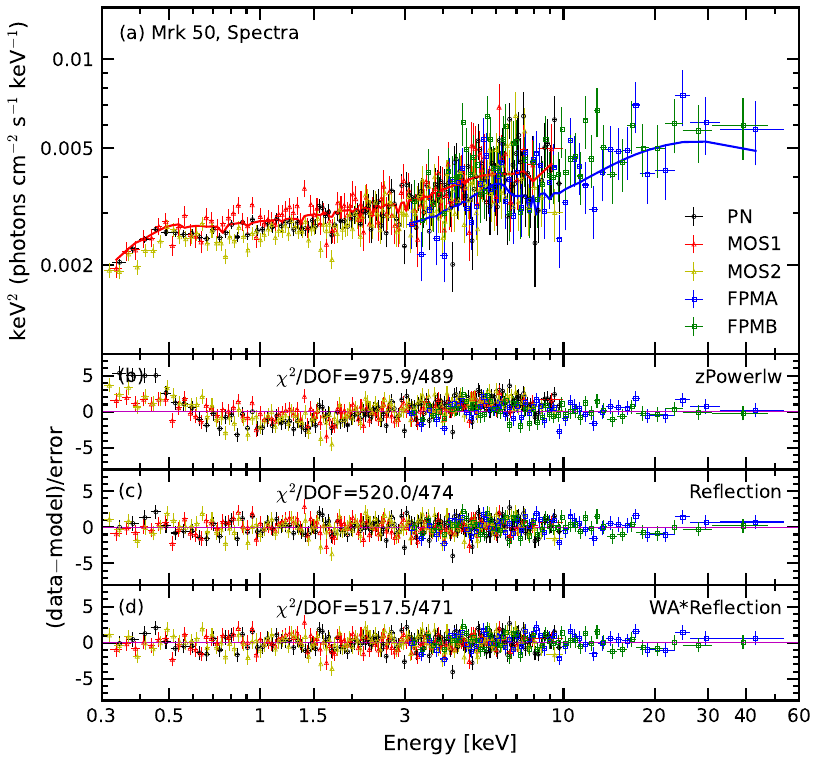}
\caption{The XMM-Newton and NuSTAR broadband spectral fit of Mrk\,50, showing the best fit with the baseline model (a). All epochs are fit simultaneously. The residuals in terms of error sigmas are plotted for (b) the phenomenological redshifted power-law model, (c) \textsc{RELXILLCP}, and (d) \textsc{RELXILLCP} plus a WA.}
\label{fig:mrk50}
\end{figure}

\subsection{Mrk\,50}
Mrk\,50 was observed by XMM-Newton once in 2009 and twice on 9 December 2010. We selected the most recent observation with the longest exposure time (0650590401, PI: L. Bassani; epoch~$a$, 22.9\,ks), and combined it with the 2022 NuSTAR observation (60061227002; epoch~$b$).

The reflection plus blackbody model improves the phenomenological fit by $\Delta \chi^2/\Delta\mathrm{DOF} = -455.6/-17$ (Figure~\ref{fig:mrk50}). Yet, the blackbody component ($kT = 0.229_{-0.069}^{+0.044}$\,keV), if removed, merely affects the fit marginally by $\Delta \chi^2/\Delta\mathrm{DOF} = +2.5/+2$, prompting us to reject this component. Further adding a WA component (Figure~\ref{fig:mrk50}d; $\Delta \chi^2/\Delta\mathrm{DOF} = -2.5/-3$) is also not well justified. This case requires a highly ionized ($\xi_\mathrm{WA} \gtrsim 10^3\,\mathrm{erg\,cm\,s^{-1}}$) WA with a column density of $N_\mathrm{H} \gtrsim 7.55 \times 10^{22}\,\mathrm{cm^{-2}}$, which seems somewhat spurious because absorption in such an environment should be weak to barely detectable. Moreover, these parameter values are close to the limits of the \textsc{XSTAR} grid points, which can produce unreliable results. Given that the goodness-of-fit is not improved by the WA, we adopt the pure reflection model (Figure~\ref{fig:mrk50}c; \textsc{CONSTANT*TBABS*RELXILLCP}) as the best fit, which has a statistic of $\chi^2/\mathrm{DOF} = 520.0/474 = 1.10$. MCMC calculations give $\theta_\mathrm{disk} = 19^\circ.1_{-3.3}^{+2.0}$ and $a_\ast = 0.992_{-0.030}^{+0.005}$.

\cite{Vasudevan et al. 2013} examined the 2009 XMM-Newton observation with both a phenomenological blackbody plus power-law ($\Gamma = 1.95\pm0.02$) model and the reflection model \textsc{PEXRAV}. With $\chi^2/\mathrm{DOF} = 902.32/879 = 1.03$, they reported a soft excess but did not detect any absorption or iron emission. Fitting the XMM-Newton spectra jointly with an observation from the Swift Burst Alert Telescope \citep[BAT;][]{Barthelmy et al. 2005} with \textsc{PEXRAV} yielded $\Gamma = 2.18 \pm 0.02$ and $R = 4.79_{-0.91}^{+0.96}$, which is slightly larger than ours, although the authors froze the abundances to solar and the inclination angle to $\cos \theta_\mathrm{disk} = 0.45$. \cite{Boissay et al. 2016} analyzed one of the 2010 XMM-Newton observations (epoch~$a$) with two Bremstrahlung components and a cutoff power-law with $\Gamma = 1.92^{+0.07}_{-0.13}$. They also calculated the reflection fraction to be $R_f = 1.75^{+0.55}_{-0.67}$ using \textsc{RELXILLLP}, which adopts a lamppost geometry \citep{Matt et al. 1991,Martocchia and Matt 1996,Reynolds and Begelman 1997,Miniutti et al. 2004} instead of our sandwich (thin coronal layers) geometry (see Section~5.6 of \citealp{Du et al. 2024} for more information on geometry selection).

\begin{figure}[t]
\centering
\includegraphics[width=1.00\linewidth]{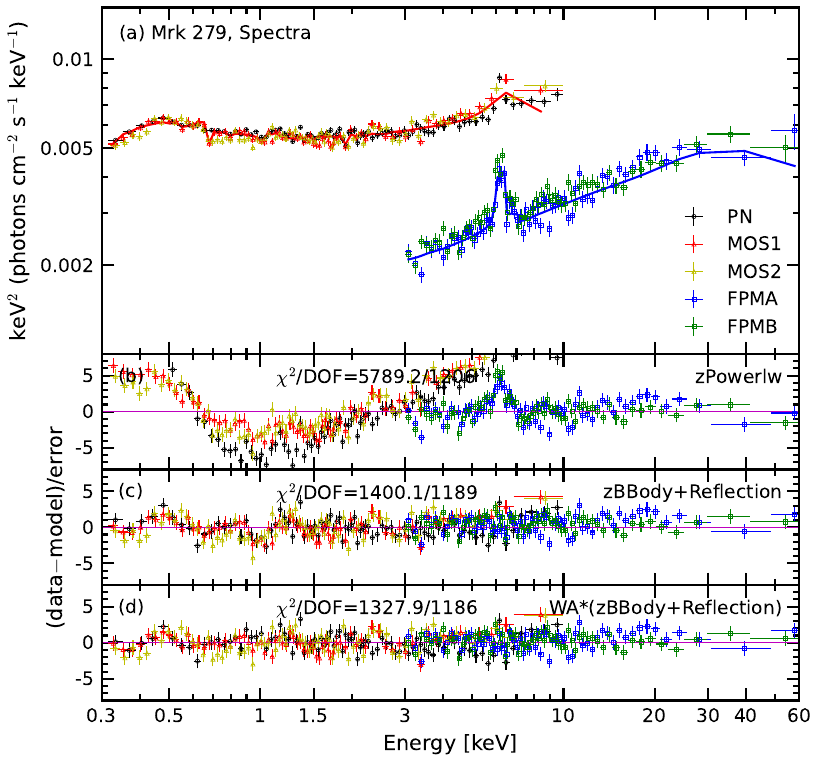}
\caption{The XMM-Newton and NuSTAR broadband spectral fit of Mrk\,279, showing the best fit with the baseline model (a). All epochs are fit simultaneously. The residuals in terms of error sigmas are plotted for (b) the phenomenological redshifted power-law model, (c) \textsc{RELXILLCP} plus a redshifted blackbody component, and (d) \textsc{RELXILLCP} plus a redshifted blackbody component and a WA. For plotting purposes, the XMM-Newton data have been rebinned to ${\rm S/N }> 25$, while the NuSTAR data have been rebinned to ${\rm S/N} > 15$.}
\label{fig:mrk279}
\end{figure}

\subsection{Mrk\,279}
The newest XMM-Newton observation of Mrk\,279 was conducted in December 2020 (0872391301; epoch~$a$, 30.5\,ks), and a long observation was completed by NuSTAR in August 2020 (60601011004; epoch~$b$). The two epochs are characterized by significantly different (factor of $\sim 2$) luminosities. The reflection plus blackbody model improves the fit by $\Delta \chi^2/\Delta\mathrm{DOF} = -4389.1/-17$ compared to the power-law model (Figure~\ref{fig:mrk279}). Adding a WA returns a best-fit model statistic of $\chi^2/\mathrm{DOF} = 1327.9/1186 = 1.12 \: (\Delta \chi^2/\Delta\mathrm{DOF} = -72.2/-3)$, while removing the blackbody component ($kT = 0.089 \pm 0.004$\,keV) from the best-fit model deteriorates the statistic by $\Delta \chi^2/\Delta\mathrm{DOF} = +123.5/+2$. Our adopted reflection modeling results in $\theta_\mathrm{disk} = 26^\circ.9_{-3.7}^{+2.5}$ and $a_\ast = 0.997_{-0.006}^{+0.001}$.

Historically, Mrk 279 has been known to be significantly variable in the X-rays from extensive observations by various missions \citep[e.g.,][]{Scott et al. 2004,Costantini et al. 2010,Ebrero et al. 2010,Yaqoob and Padmanabhan 2004,Jiang et al. 2019,Igo et al. 2020,Ursini et al. 2020}. \cite{Akhila et al. 2024} reported long-term X-ray analysis of this source during 2018--2020, including both epochs~$a$ and $b$. They fit epoch~$a$ with a blackbody component, a thermal Comptonization component \citep[\textsc{THCOMP};][]{Zdziarski et al. 2020}, a Gaussian line at $6.41 \pm 0.04 \,\rm keV$, and a Compton scattering power-law \citep[\textsc{SIMPL};][]{Steiner et al. 2009} with $\Gamma = 1.48^{+0.16}_{-0.21}$. Another fit for epoch~$b$ without the blackbody yielded $\Gamma = 1.62 \pm 0.01$. \cite{Ballantyne et al. 2024} studied three XMM-Newton observations from 2005 with the reflection model \textsc{REXCOR} \citep{Xiang et al. 2022} with lamppost geometry at a fixed inclination of $30^\circ$, and obtained $a_\ast = 0.99$ and the lamppost coronal height varying from $20\,r_g$ to $5\,r_g$.

\begin{figure}[t]
\centering
\includegraphics[width=1.00\linewidth]{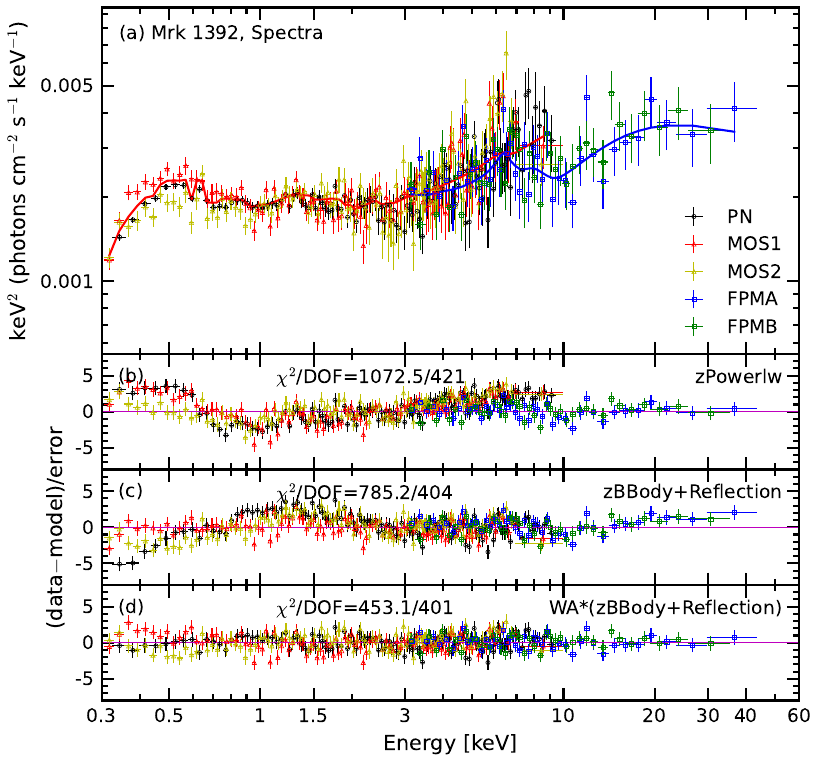}
\caption{The XMM-Newton and NuSTAR broadband spectral fit of Mrk\,1392, showing the best fit with the baseline model (a). All epochs are fit simultaneously. The residuals in terms of error sigmas are plotted for (b) the phenomenological redshifted power-law model, (c) \textsc{RELXILLCP} plus a redshifted blackbody component, and (d) \textsc{RELXILLCP} plus a redshifted blackbody component and a WA.}
\label{fig:mrk3192}
\end{figure}

\subsection{Mrk\,1392}
Mrk\,1392 was observed by XMM-Newton on 20 January 2018 (0795670101, PI: G. Lansbury; epoch~$a$, 38\,ks) and by NuSTAR on 25 January 2018 (60160605002; epoch~$b$). As with MCG\,+04$-$22$-$042, even the non-simultaneous NuSTAR data can be useful for broadband modeling of the hard X-rays. Relative to the power-law model, the \textsc{RELXILLCP} plus blackbody model improves the phenomenological fit by $\Delta \chi^2/\Delta\mathrm{DOF} = -287.3/-17$ (Figure~\ref{fig:mrk3192}). Adding a WA returns the best-fit model statistic of $\chi^2/\mathrm{DOF} = 453.1/401 = 1.13 \: (\Delta \chi^2/\Delta\mathrm{DOF} = -332.1/-3)$. If removing the blackbody component ($kT = 0.040_{-0.008}^{+0.005}$\,keV) from the best-fit model, the statistic deteriorates by $\Delta \chi^2/\Delta\mathrm{DOF} = +9.6/+2$. Our preferred reflection model results in $\theta_\mathrm{disk} = 25^\circ.5_{-9.9}^{+6.7}$ and $a_\ast = 0.942_{-0.093}^{+0.038}$.

\cite{Esparza-Arredondo et al. 2020} fit the NuSTAR spectra of epoch~$b$ using a single power-law with a partial covering absorber to account for Galactic absorption. Achieving $\chi^2/\mathrm{DOF} = 1.12$, they reported $\Gamma = 1.84^{+0.05}_{-0.04}$, consistent with our results. \cite{Esparza-Arredondo et al. 2021} repeated the analysis using alternative models that account for reflection features caused by a distant reflecting torus. Their values of $\Gamma$ are roughly compatible with previous results, but their smooth torus model \citep[borus02;][]{Balokovic et al. 2018} returned an inclination of $87.0^\circ$, while their clumpy model \citep[UXClumpy;][]{Buchner et al. 2019} gave a markedly different inclination of $1.0^\circ$. Another difference between our analysis and theirs regards the interpretation of Galactic absorption. In their first paper, the measurement for the column density was $\log (N_\mathrm{H}/\rm cm^{-2}) = 24.63^{+0.16}_{-0.18}$; their subsequent work obtained a similar large value of $N_\mathrm{H}$. We find substantially different results. Instead of fitting the Galactic absorption with a partial covering absorber, we use the \textsc{TBABS} model with a column density fixed to $\log (N_\mathrm{H}/\rm cm^{-2}) = 20.58$ from the Leiden/Argentine/Bonn Survey. We also find a WA with $\log (N_\mathrm{H}/\rm cm^{-2}) \approx 21$.

\begin{figure}[t]
\centering
\includegraphics[width=1.00\linewidth]{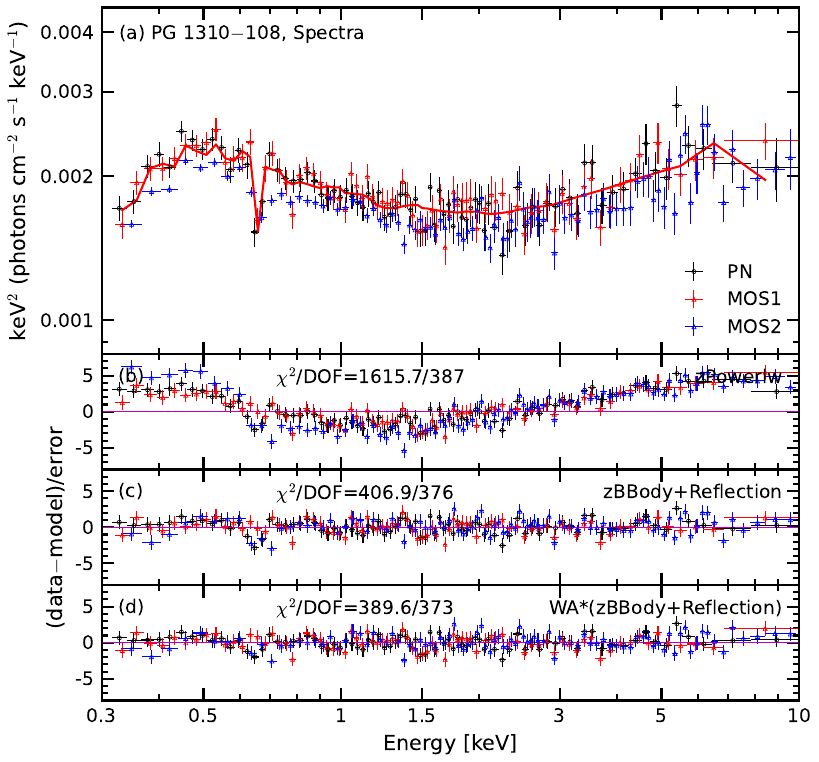}
\caption{The XMM-Newton and NuSTAR broadband spectral fit of PG\,1310$-$108, showing the best fit with the baseline model (a). All epochs are fit simultaneously. The residuals in terms of error sigmas are plotted for (b) the phenomenological redshifted power-law model, (c) \textsc{RELXILLCP} plus a redshifted blackbody component, and (d) \textsc{RELXILLCP} plus a redshifted blackbody component and a WA. For plotting purposes, the XMM-Newton data have been rebinned to ${\rm S/N} > 10$.}
\label{fig:pg1310}
\end{figure}

\subsection{PG\,1310$-$108}
PG\,1310$-$108 was observed by XMM-Newton once in 2018 for 21\,ks (0801891601, PI: S. Kaspi). Lacking observations at higher energy, the sole EPIC spectrum is adopted as epoch~$a$ in our analysis. All disk- and corona-related parameters are free to vary. The reflection plus blackbody model improves the power-law fit by $\Delta \chi^2/\Delta\mathrm{DOF} = -1,208.8/-11$ (Figure~\ref{fig:pg1310}). Including a WA yields a best-fit model statistic of $\chi^2/\mathrm{DOF} = 389.6/373 = 1.04 \: (\Delta \chi^2/\Delta\mathrm{DOF} = -17.3/-3)$, while excluding the blackbody component with $kT = 0.089_{-0.010}^{+0.007}$\,keV gives $\Delta \chi^2/\Delta\mathrm{DOF} = +22.5/+2$, which shows the necessity of this component. Our preferred reflection model results in $\theta_\mathrm{disk} = 33^\circ.5_{-22.4}^{+20.8}$ and $a_\ast = 0.903_{-0.403}^{+0.091}$. The accretion disk inclination is $\sim 10^\circ$ smaller than the BLR inclination of this source, although lying within the error range. The uncertainties are relatively large because only a single epoch of XMM-Newton observation is available. The lack of higher energy data lost the important information regarding the Compton hump, which is crucial for constraining the reflection fraction, and thus, the systematic errors as well as the statistical errors for the reflection model. Nevertheless, we can obtain an upper limit for the reflection fraction of $R_f = 2.02$, with the 50\% value on the MCMC chain equal to $R_f = 0.88$, which implies that $0.88/(1+0.88) = 47\%$ of the spectrum arises from reflection. The results from the reflection model should be reasonably reliable. Future observations with broader energy coverage and more (simultaneous) spectral epochs will improve the constraints.

\begin{figure}[t]
\centering
\includegraphics[width=1.00\linewidth]{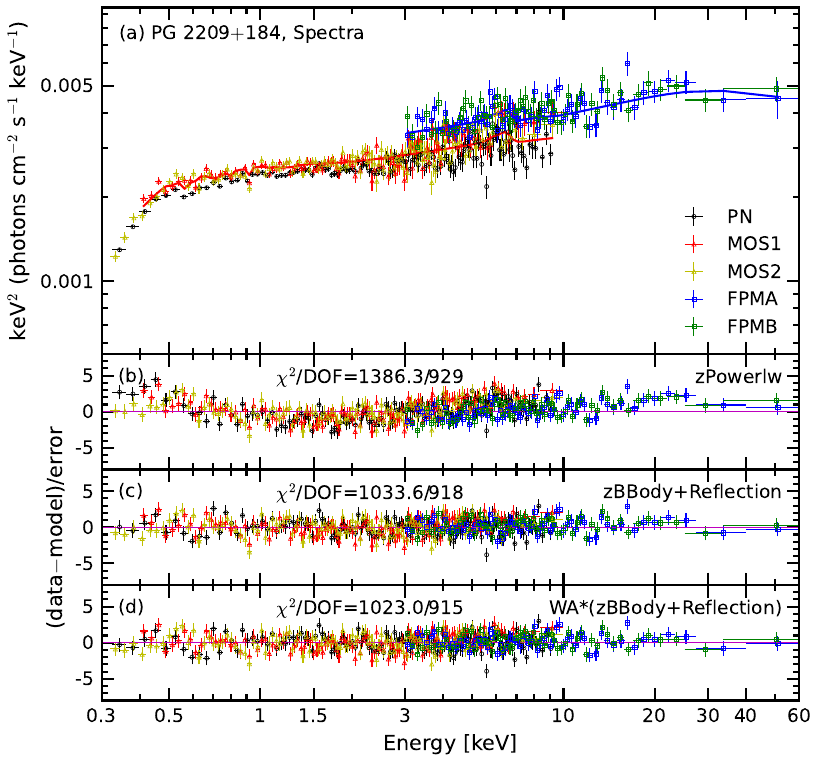}
\caption{The XMM-Newton and NuSTAR broadband spectral fit of PG\,2209+184, showing the best fit with the baseline model (a). All epochs are fit simultaneously. The residuals in terms of error sigmas are plotted for (b) the phenomenological redshifted power-law model, (c) \textsc{RELXILLCP} plus a redshifted blackbody component, and (d) \textsc{RELXILLCP} plus a redshifted blackbody component and a WA. For plotting purposes, data have been rebinned to ${\rm S/N} > 10$.}
\label{fig:pg2209}
\end{figure}

\subsection{PG\,2209+184}
PG\,2209+184 was observed by XMM-Newton on 22 November 2017 (0795620201, PI: K. Koljonen; 54.9\,ks) and on 30 November 2017, and then again by NuSTAR on 23 November 2017 (60301015002; 101.9\,ks). The 22 November observation ended on midnight, only less than 2 hours earlier than the beginning of the NuSTAR observation. Considering that strong variability in X-rays has not been reported for this source, we assume the two epochs to be simultaneous and designate them both as epoch~$a$. In this case, the corona-related parameters are tied between the two epochs. 

As shown in Figure~\ref{fig:pg2209}, the reflection plus blackbody model improves the power-law fit by $\Delta \chi^2/\Delta\mathrm{DOF} = -352.7/-11$, adding a WA gives $\chi^2/\mathrm{DOF} = 1,023.0/915 = 1.12 \: (\Delta \chi^2/\Delta\mathrm{DOF} = -10.6/-3)$, and omitting the blackbody component with $kT = 0.086_{-0.017}^{+0.015}$\,keV affects the model by $\Delta \chi^2/\Delta\mathrm{DOF} = +25/+2$. Our reflection modeling results in $\theta_\mathrm{disk} = 35^\circ.6_{-25.9}^{+31.7}$ and $a_\ast = 0.989_{-0.079}^{+0.009}$. The constraints are rather loose, even though an epoch of simultaneous data were fit, on account of the low reflection fraction of $0.21 < R_f < 0.72$, for which at most $0.72/(1+0.72) = 42\%$ of the spectrum originates from the ionized reflector.

We also test the validity of combining the two observations into one epoch by comparing the results with those from fitting the two observations separately. We untie the corona-related parameters between the two observations and produce the fit and MCMC calculations. In the two-epoch fit, the accretion disk inclination is $\theta_\mathrm{disk} = 44.2_{-4.5}^{+7.9}$ and the spin is $a_\ast = 0.980_{-0.058}^{+0.018}$, which are consistent with the results from the one-epoch fit. The reflection fractions for the two epochs are $0.27 < R_f < 0.44$ and $0.10 < R_f < 0.24$, respectively. Their photon indices are exactly the same and equal to the index derived when fit being combined ($\approx 1.90$), and other parameters are also consistent. These evidence support our choice of combining the two observations into one epoch. However, treating the two observations separately with an additional group of seven free corona-related parameters makes the fit harder to converge, for which a MCMC chain of three times the length of the one-epoch fit is applied here. Moreover, the fit statistic is slightly worse, with $\chi^2/\mathrm{DOF} = 1018.3/908 = 1.121$ comparing to 1.118 from the one-epoch fit ($\Delta \chi^2/\Delta\mathrm{DOF} = -4.7/-7$). The results from the single epoch scenario are thus adopted.

\begin{figure}[t]
\centering
\includegraphics[width=1.00\linewidth]{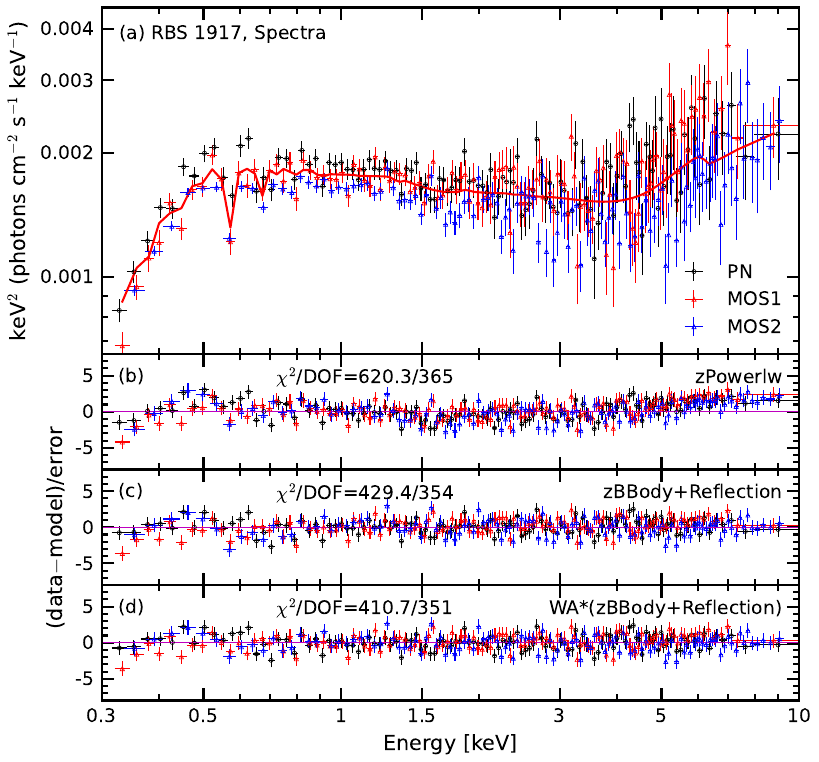}
\caption{The XMM-Newton and NuSTAR broadband spectral fit of RBS\,1917, showing the best fit with the baseline model (a). All epochs are fit simultaneously, and the residuals in terms of error sigmas are plotted for (b) the phenomenological redshifted power-law model, (c) \textsc{RELXILLCP} plus a redshifted blackbody component, and (d) \textsc{RELXILLCP} plus a redshifted blackbody component and a WA.}
\label{fig:rbs1917}
\end{figure}

\subsection{RBS\,1917}
RBS\,1917 was observed by XMM-Newton twice in 2015. We select the November observation with 35.8\,ks duration (0762871101, PI: N. Okabe). As with PG\,1310$-$108, the source lacks higher energy observations. The reflection plus blackbody model improves the power-law model by $\Delta \chi^2/\Delta\mathrm{DOF} = -190.9/-11$, whereas including a WA reduces the fit statistic to optimal $\chi^2/\mathrm{DOF} = 410.7/351 = 1.17 \: (\Delta \chi^2/\Delta\mathrm{DOF} = -18.7/-3)$ (Figure~\ref{fig:rbs1917}). The blackbody component ($kT = 0.248_{-0.031}^{+0.009}$\,keV) is necessary because removing it alters the fit by $\Delta \chi^2/\Delta\mathrm{DOF} = +28.3/+2$. Our reflection modeling results in $\theta_\mathrm{disk} = 17^\circ.2_{-1.1}^{+4.5}$ and $a_\ast = 0.977_{-0.003}^{+0.012}$. With only one epoch of XMM-Newton data available, the model values, especially those related to broadband or hard X-ray spectral features, are not tightly constrained during the fit. We can derive only rather unlikely lower limits on the reflection fraction ($R_f > 9.87$) and coronal temperature ($kT_e > 380$\,keV). Nevertheless, the best-fit inner accretion disk inclination of $\sim 17^\circ$ agrees reasonably closely with the reported BLR inclination of $\sim 20^\circ$.

\begin{figure}[t]
\centering
\includegraphics[width=1.00\linewidth]{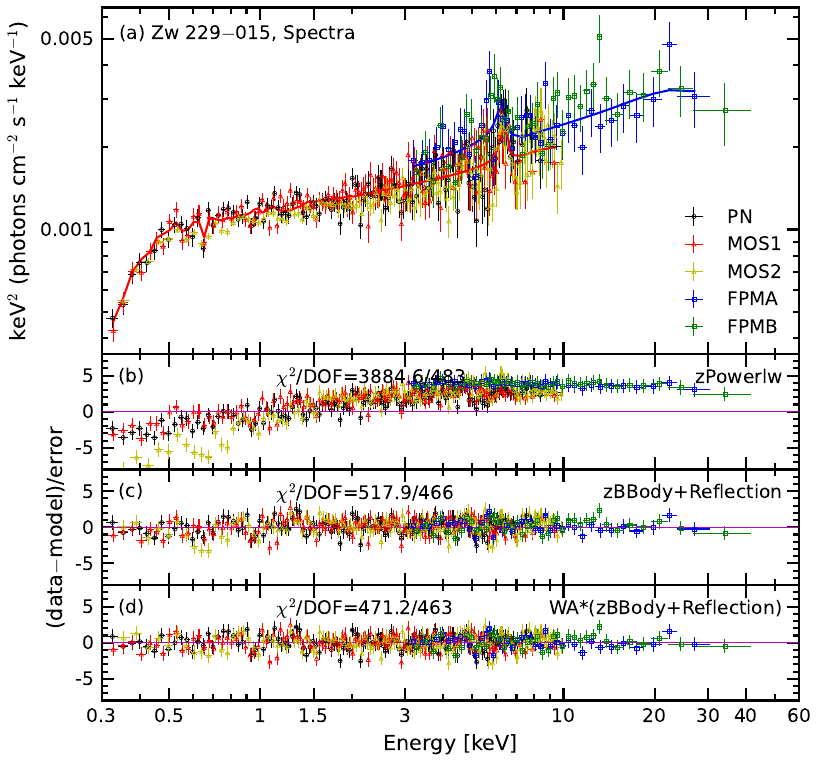}
\caption{The XMM-Newton and NuSTAR broadband spectral fit of Zw\,229$-$015, showing the best fit with the baseline model (a). All epochs are fit simultaneously. The residuals in terms of error sigmas are plotted for (b) the phenomenological redshifted power-law model, (c) \textsc{RELXILLCP} plus a redshifted blackbody component, and (d) \textsc{RELXILLCP} plus a redshifted blackbody component and a WA.}
\label{fig:zw229}
\end{figure}

\subsection{Zw\,229$-$015}
Zw\,229$-$015 was observed by XMM-Newton in 2011 (0672530301, PI: L. Gallo; epoch~$a$, 29.1\,ks) and then again by NuSTAR in 2018 (60160705002; epoch~$b$). The two observations are not simultaneous, and thus we treat them as separate epochs and use independent corona-related parameters to describe them respectively. As illustrated in Figure~\ref{fig:zw229}, the reflection plus blackbody model improves the power-law model by $\Delta \chi^2/\Delta\mathrm{DOF} = -3,366.7/-17$. Introducing a WA gives a best-fit statistic of $\chi^2/\mathrm{DOF} = 471.2/463 = 1.018 \: (\Delta \chi^2/\Delta\mathrm{DOF} = -46.7/-3)$, while removing the blackbody component ($kT = 0.103 \pm 0.001$\,keV) degrades the fit by $\Delta \chi^2/\Delta\mathrm{DOF} = +18/+2$. Our reflection modeling indicates $\theta_\mathrm{disk} = 26^\circ.2_{-0.5}^{+2.5}$ and $a_\ast = 0.993_{-0.059}^{+0.005}$.

\cite{Adegoke et al. 2017} studied the XMM-Newton spectra and concluded that thermal Comptonization and blurred reflection were equally effective in explaining the data, offering physical insights to the soft excess by comparing the multicolor disk blackbody and smeared wind absorption models. Their Comptonization \citep[\textsc{COMPTT};][]{Titarchuk 1994} plus redshifted power-law model gave a plasma temperature of $0.632 \pm 0.232$\,keV and $\Gamma=1.52 \pm 0.10$. Fixing the inclination to $30^\circ$, they used reflection calculations of \textsc{REFLIONX} \citep{Ross and Fabian 2005} and the blurring kernel \textsc{LAOR} \citep{Laor 1991} to derive a photon index of $\Gamma = 1.52 \pm 0.082$, which is low compared to our value of $\Gamma \approx 1.8$, and a low iron abundance of $0.469 \pm 0.095\,A_\odot$ and yet a high ionization state of $\xi \approx 2300\,\rm erg\,cm\,s^{-1}$. Their fit returned $\chi^2/\mathrm{DOF} = 1436/1404 = 1.023$, which is merely $0.5\%$ bigger than our best-fit model. The discrepancy in the photon index is likely due to the different reflection models used. They also reported the blur index (uniform emissivity index) at $4.22 \pm 1.45$ that is smaller than both of our indices, and the inner disk radius at $3.999 \pm 0.491$ that is larger than our value given $a_\ast = 0.993$. \cite{Tripathi et al. 2019} conducted detailed analysis of this source using data from both epochs, as well as additional data from Swift XRT/BAT and Suzaku. They fit the spectra with an ionized partial covering model, \textsc{RELXILL}, and soft Comptonization, respectively, but were also unable to select the best model statistically. They fixed $a_\ast = 0.998$ and $\theta_\mathrm{disk} = 45^\circ$ for the reflection model. In a first attempt with XMM-Newton and Suzaku data, they obtained $\log (\xi/\mathrm{erg\,cm\,s^{-1}}) = 2.56 \pm 0.15$ and $R_f \approx 0.4$ with a reduced $C$-stat of $411/355 = 1.16$, parameter values that are consistent with ours. Multi-epoch fits that include NuSTAR data resulted in a reduced $C$-stat of $561/459 = 1.22$ with $\log (\xi / \mathrm{erg\,cm\,s^{-1}}) = 2.64_{-0.17}^{+0.07}$. Their photon indices for epoch $a$ and $b$ are $\Gamma \approx 1.88$ and $\Gamma \approx 1.93$, respectively. Our photon indices are larger than that from \cite{Adegoke et al. 2017} by $\Delta \Gamma \approx 0.2$ and smaller than that from \cite{Tripathi et al. 2019} by $\Delta \Gamma \approx 0.1$.

\bibliographystyle{aasjournal}

\end{document}